\documentclass[11pt, a4paper]{article}
\usepackage{amsmath}

\newtheorem{theorem}{Theorem}

\begin{document}

\title{Differential-geometric approach to the integrability of hydrodynamic chains: the Haantjes tensor}
\author{E.V. Ferapontov  and D.G. Marshall}
    \date{}
    \maketitle
    \vspace{-7mm}
\begin{center}
Department of Mathematical Sciences \\ Loughborough University \\
Loughborough, Leicestershire LE11 3TU \\ United Kingdom \\[2ex]
e-mails: \\[1ex] \texttt{E.V.Ferapontov@lboro.ac.uk}\\
\texttt{D.G.Marshall@lboro.ac.uk}
\end{center}

\bigskip

\begin{abstract}
The integrability of an $m$-component system of hydrodynamic type,
${\bf u}_t=V({\bf u}){\bf u}_x$,
 by the generalized hodograph method requires the diagonalizability of the $m \times m$ matrix $V({\bf u})$. This condition is known to be  equivalent to the vanishing of the corresponding Haantjes tensor. 
We generalize this approach  to hydrodynamic chains ---  infinite-component systems of hydrodynamic type for which the $\infty \times \infty$ matrix $V({\bf u})$ is `sufficiently sparse'. For such systems the Haantjes tensor is well-defined, and the calculation of its components involves finite summations only. We illustrate our approach  by  classifying broad classes of conservative  and Hamiltonian hydrodynamic chains with the zero Haantjes tensor. 
We prove that the vanishing of the Haantjes tensor  is a necessary condition for a hydrodynamic chain to possess an infinity of semi-Hamiltonian hydrodynamic reductions, thus providing an easy-to-verify necessary condition for the integrability.  

\bigskip


\bigskip

\end{abstract}

\newpage

\section{Introduction}

Hydrodynamic chains are  quasilinear first order PDEs of the form
\begin{equation}
{\bf u}_t=V({\bf u}) {\bf u}_x
\label{chain}
\end{equation}
where ${\bf u}=(u^1, u^2, ...)^t$ is an infinite-component column vector and $V({\bf u})$ is an $\infty \times \infty$ matrix. 
A classical example is the Benney chain (or Benney's moment equations),  
\begin{equation}
u^n_t=u^{n+1}_x+(n-1)u^{n-1}u^1_x, 
\label{Benney}
\end{equation}
$n=1, 2, ...$, which was  derived  in  \cite{Benney} from the  equations for long nonlinear waves on a free surface. It was observed later in \cite{Zakharov1, Gibb81} that the same system  results from a  kinetic Vlasov  equation.  The system (\ref{Benney})  has been thoroughly investigated in the subsequent publications
\cite{Miura, Kup1, Kup2, Lebedev, Zakharov1, Zakharov2, T1, Gibb81} where, in particular,  its Hamiltonian and integrability aspects were uncovered. Hydrodynamic reductions of the chain (\ref{Benney}) were studied in \cite{Gibb94, GibTsa96, GibTsa99, Krichever, Ma3}. 
Various  deformations of Benney's equations are known. These include the modified Benney
chain,
\begin{equation}
u^n_t=u^{n+1}_x+u^1u^n_x+(n-1)u^nu^1_x, 
\label{mod}
\end{equation}
obtained in \cite{Kup4}  as a quasiclassical limit of the modified KP hierarchy. 
Its  two-parameter deformation,
\begin{equation}
u^n_t=u^{n+1}_x+u^1u^n_x+(a(n-1)+b)u^nu^1_x, 
\label{Kup}
\end{equation}
was constructed in \cite{Kup3} along with further examples of Hamiltonian integrable chains possessing  complete systems of commuting integrals. Another deformation scheme, based on the $R$-matrix approach, was proposed in \cite { Bla1}, see also \cite{ Ma1, Ma2}. The specialization of the chain (\ref{Kup}) corresponding  to $a=0, \ b=-1$, 
\begin{equation}
u^n_t=u^{n+1}_x+u^1u^n_x-u^nu^1_x, 
\label{uni}
\end{equation}
naturally appears in the theory of finite-gap solutions of integrable hierarchies of the KdV type
\cite{Alber, Mikhalev}. Reductions of the chain (\ref{uni}), both hydrodynamic and differential,  were extensively investigated in \cite{Pavlov1, Shabat1, Shabat2}. The case $a=1, \ b=2$ arises from the kinetic model  for rarefied bubbly flows \cite{K}, see also \cite{T2} for an alternative representation of this chain.

A broad  class of  new examples was found  in \cite{Pavlov2}, see also \cite{Buk}, based on the symmetry approach. These  papers provide a  classification of  conservative chains of the form
\begin{equation}
\begin{array}{c}
u^1_t=u^2_x, ~~~ u^2_t=g(u^1, u^2, u^3)_x, ~~~ u^3_t=h(u^1, u^2, u^3, u^4)_x,  ...,
\end{array}
\label{max}
\end{equation}
which are embedded into a commutative hierarchy of Egorov's type. It was observed  that  the function $g(u^1, u^2, u^3)$ uniquely determines all other equations of the chain (\ref{max}), as well as the whole associated hierarchy. Moreover, the function $g(u^1, u^2, u^3)$ was shown to satisfy an over-determined involutive system of third order PDEs whose general solution was expressed  in terms  of theta functions and solutions to the Chazy equation.

\medskip

Our approach to the integrability of hydrodynamic chains is motivated by the theory of finite-component
systems of hydrodynamic type, that is, equations of the form (\ref{chain}) where ${\bf u}=(u^1, u^2, ..., u^m)^t$ is an $m$-component column vector and $V({\bf u})=v^i_j({\bf u})$ is  an $m \times m$ matrix.
Explicitly, one has
\begin{equation}
u^i_t=v^i_j({\bf u}) u^j_x, ~~~ i, j=1,..., m;
\label{1d}
\end{equation}
here and below we adopt the standard summation convention over the repeated upper and lower indices. Such systems naturally occur in applications in gas dynamics, fluid mechanics, chemical kinetics, Whitham averaging procedure, differential geometry and topological field theory. We refer to \cite{Tsarev, Dub,  Dubr} for a further discussion and references. In what follows we consider the strictly hyperbolic case when the eigenvalues of the matrix $v^i_j$, called the characteristic speeds of  the system  (\ref{1d}), are real and distinct. 

It is known that many particularly important systems of the form (\ref{1d})  are diagonalizable, that is, reducible to the Riemann invariant form
\begin{equation}
R^i_t=\lambda^i({\bf R}) R^i_x
\label{1R}
\end{equation} 
where the  characteristic speeds $\lambda^i(\bf R)$ satisfy the so-called semi-Hamiltonian property
\cite{Tsarev}, 
\begin{equation}
\partial_k\left(\frac{\partial_j \lambda^i}{\lambda^j-\lambda^i} \right)= \partial_j\left(\frac{\partial_k \lambda^i}{\lambda^k-\lambda^i} \right), 
\label{semi}
\end{equation}
$\partial_k=\partial/\partial R^k$, \  $i\ne j\ne k$. We emphasize that the semi-Hamiltonian property (\ref{semi}) is usually automatically satisfied for  diagonalizable systems of the `physical' origin. For instance, a conservative diagonalizable system is necessarily semi-Hamiltonian (see  Appendix 1 for more details). Such systems  possess infinitely many
conservation laws and commuting flows of hydrodynamic type and can be linearized by the generalized hodograph method \cite{Tsarev}.  Their analytic, differential-geometric and Hamiltonian aspects are 
well-understood by now. 

Remarkably, there exists an efficient tensor criterion of  the diagonalizability   which does not require the actual computation of  eigenvalues and  eigenvectors of the matrix $v^i_j$.
Let us first calculate the Nijenhuis tensor of the matrix $v^i_j$,
\begin{equation}
N^i_{jk}=v^p_j\partial_{u^p}v^i_k-v^p_k\partial_{u^p}v^i_j-v^i_p(\partial_{u^j}v^p_k-\partial_{u^k}v^p_j),
\label{N}
\end{equation}
 and introduce the Haantjes tensor
\begin{equation}
H^i_{jk}=N^i_{pr}v^p_jv^r_k-N^p_{jr}v^i_pv^r_k-N^p_{rk}v^i_pv^r_j+N^p_{jk}v^i_rv^r_p.
\label{H}
\end{equation}
For strictly hyperbolic systems  the condition of diagonalizability is given by the following theorem.
\begin{theorem} \cite{Haantjes}
 A  hydrodynamic type system  with  mutually distinct characteristic speeds is diagonalizable if and only if the corresponding Haantjes tensor (\ref{H}) is identically zero. 
\end{theorem}
Since  components of the Haantjes tensor can be calculated (using  computer algebra) in any  coordinate system, this provides an efficient diagonalizability criterion. 

\medskip

Our main observation is that both tensors (\ref{N}) and (\ref{H}) make perfect sense for infinite matrices which are `sufficiently sparse'. To be more rigorous, let us give the following 

\medskip

\noindent {\bf Definition 1.} {\it
An infinite matrix $V({\bf u})$ is said to belong to the class $C$ (chain class) if it satisfies the following two natural properties:

\noindent (a) each row of $V({\bf u})$ contains {\it finitely many} nonzero elements;

\noindent (b) each matrix element of $V({\bf u})$ depends on {\it finitely many} variables $u^i$.}

\medskip

\noindent Notice that the chains (\ref{Benney}) - (\ref{max}) clearly belong to the class $C$. For matrices from the class $C$ all contractions in the expressions (\ref{N}) and (\ref{H}) reduce to {\it finite} summations so that each particular component $H^i_{jk}$  is a well-defined object which can be effectively computed. Moreover, for a fixed value of an upper index $i$ there exist only {\it finitely many} non-zero components $H^i_{jk}$. 

We propose the following 

\medskip

\noindent {\bf Definition 2.} {\it
A hydrodynamic chain  from the class $C$ is said to be {\it diagonalizable} if all components of the corresponding Haantjes tensor (\ref{H}) are zero}. 

\medskip

\noindent We point out that  the chains (\ref{Benney}) - (\ref{uni}) are diagonalizable in this sense. As we prove in Sect. 5, the vanishing of the Haantjes tensor is a necessary (and in many particularly important cases --- sufficient) condition for a hydrodynamic chain to possess an infinity of finite-component diagonalizable hydrodynamic reductions, thus justifying the above definition.
The main advantage of our approach is its `intrinsic' character:  it does not require the knowledge of any `extrinsic' objects such as commuting flows, Hamiltonian structures, Lax pairs, etc. 

The vanishing of the Haantjes tensor turns out to be an efficient classification criterion.
As an elementary example let us consider  the chain
$$
u^n_t=u^{n+1}_x+u^1u^n_x+c_nu^nu^1_x
$$
where $c_n=const$. One can readily verify that the vanishing of the Haantjes tensor implies the recurrence relation $c_{n+2}=2c_{n+1}-c_n$. Setting $c_1=b, \ c_2=a+b$ we recover the integrable chain (\ref{Kup}).  

Based on the same criterion,  in Sect. 2  we classify diagonalizable chains of the type (\ref{max}). It turns out that the conditions  $H^1_{jk}=0$ are already sufficiently restrictive and imply an over-determined system expressing all second order partial derivatives of  the function $h$ in terms of $g$, see  (\ref{h}). The consistency conditions of these equations lead to a closed-form involutive system  expressing all third order partial derivatives of $g$ in terms of its lower order derivatives, see (\ref{g}) (we have extensively used computer algebra to calculate the Haantjes tensor and to verify the involutivity by calculating the compatibility conditions). %We emphasize that exactly the same system  for $g$  was derived in \cite{Pavlov2} using the symmetry approach, as well as in \cite{Fer7} based on the method of hydrodynamic reductions. 
%Thus, for conservative chains  (\ref{max}) the condition of diagonalizability is equivalent to the integrability. 
The requirement of the vanishing of other components $H^i_{jk}, \  i\geq 2$,  imposes no additional constraints on $h$ and $g$: these conditions  reconstruct  the remaining equations of the chain (\ref{max}).
For instance, the conditions $H^2_{jk}=0$ specify the right hand side of the fourth equation  $u^4_t=...$, etc.

The characterization of diagonalizable chains of a more general form,
\begin{equation}
\begin{array}{c}
u^1_t=f(u^1, u^2)_x, ~~~
u^2_t=g(u^1, u^2, u^3)_x, ~~~ u^3_t=h(u^1, u^2, u^3, u^4)_x, ...,
\end{array}
\label{max1}
\end{equation}
etc, is proposed  in Sect. 3. As in the previous example,  the conditions  $H^1_{jk}=0$ lead to  expressions for all second order partial derivatives of $h$ in terms of $g$ and $f$. The consistency conditions   of these equations result in an  involutive  system expressing all third order partial derivatives of $g$ and $f$ in terms of lower order derivatives thereof. 

In Sect. 4 we  classify  diagonalizable Hamiltonian chains of the form
\begin{equation}
{\bf u}_t=\left (B\frac{d}{dx}+\frac{d}{dx}B^t \right )\frac{\partial h}{\partial {\bf u}},
\label{Ham}
\end{equation}
here $B^{ij}=(i-1)u_{i+j-2}$ and $h(u^1, u^2, u^3)$ is a Hamiltonian density.
The Benney chain (\ref{Benney}) corresponds to  $h=(u^3+(u^1)^2)/2$, see \cite{Kup2}. We have found a broad family of new integrable Hamiltonian densities, 
%in particular, $h=(u^3+P(u^1, u^2))^{1/3}$ where $P$ is a  cubic polynomial, 
thus extending the results of  \cite{Kup3}.

In Sect. 5 we prove that the condition of diagonalizability is necessary for the existence of `sufficiently many' hydrodynamic reductions. Recall that an $m$-component hydrodynamic reduction of an infinite chain is specified by parametric equations
$$
u^1=u^1(R^1, \dots, R^m), ~~~ u^2=u^2(R^1, \dots, R^m), ~~~ u^3=u^3(R^1, \dots, R^m), ...,
$$
etc, where the  Riemann invariants $R^1, \dots, R^m$ solve a diagonal system (\ref{1R}) whose characteristic speeds satisfy the semi-Hamiltonian property (\ref{semi}). It is required that all equations of the chain are satisfied identically modulo  (\ref{1R}). Thus, an infinite chain  reduces to a system with finitely many dependent variables. It was demonstrated in \cite{GibTsa96}
that the Benney chain (\ref{Benney}) possesses infinitely many $m$-component reductions of this type
parametrized by $m$ arbitrary functions of a single variable. The same is true for other chains (\ref{Benney}) - (\ref{max}). Based on these and other examples we propose the following 

\medskip

\noindent {\bf Definition 3.} {\it
A hydrodynamic chain  from the class $C$ is said to be {\it integrable} if, for any $m$,  it possesses infinitely many $m$-component semi-Hamiltonian reductions parametrised by $m$ arbitrary functions of a single variable.}

\medskip

\noindent {\bf Remark.} It was observed in  \cite{Kup6, Pavlov4} that  Hamiltonian chains (\ref{Ham})   possess  $m$-component reductions for {\it any} Hamiltonian density $h$, even in the non-integrable case. The crucial difference is that for integrable chains these reductions are semi-Hamiltonian and depend on $m$ arbitrary functions of a single variable, while in the non-integrable situation for any $m$ there exists a {\it unique} $m$-component reduction which is {\it not}  diagonalizable.
  
\medskip

In Sect. 5 we prove our main result:

\begin{theorem}
The vanishing of the Haantjes tensor $H$ is a necessary condition for the integrability of  hydrodynamic chains from the class $C$.
\label{T}
 \end{theorem}
 
 \medskip
 
If the spectrum of the infinite matrix $V$ is simple, that is,  for a generic $\lambda$ there exists a unique formal eigenvector $\xi$ such that $V\xi=\lambda \xi$ (we point out that all of the above examples satisfy this property), one has the following stronger result:

\begin{theorem}
In the  simple spectrum case the vanishing of the Haantjes tensor $H$ is  necessary and sufficient  for the existence of two-component reductions parametrized by two arbitrary functions of a single variable.
\label{T1}
 \end{theorem}
 
 \medskip

Theorem 2 provides an easy-to-verify {\it necessary} condition for testing the integrability of hydrodynamic chains. We emphasize that the vanishing of the Haantjes tensor  is not sufficient for the integrability in general: one can  construct examples  of diagonalizable chains which possess infinitely many diagonal  reductions {\it none} of which are semi-Hamiltonian (see Sect. 5). To eliminate these pathological cases let us recall that for finite-component systems (\ref{1d}) there exists a  tensor object which is responsible for the semi-Hamiltonian property \cite{Pavlov3}. This is a $(1, 3)$-tensor $P^s_{kij}$ (see  Appendix 1 for explicit formulas in terms of the matrix $v^i_j$). Similarly to the Haantjes tensor $H$, the tensor $P$ is well-defined for hydrodynamic chains from the class $C$. We conclude this introduction by formulating the following

\medskip

\noindent {\bf Conjecture.} {\it
The vanishing of both tensors $H$ and $P$ is  necessary and sufficient for the integrability of   hydrodynamic chains from the class $C$.}

\medskip

\noindent The necessity of this conjecture (that is, the statement that the integrability implies the vanishing of both $H$ and $P$) is a relatively simple fact, see Sect. 5 for a proof. The sufficiency is a far more delicate property which we were not able to establish in general. 
We point out that the vanishing of $H$ alone (in fact, the vanishing of the very few first components of $H$), is already sufficiently restrictive and implies the integrability in many particularly important cases (e.g. for conservative chains, Hamiltonian chains, etc). 
Recall that, for finite-component systems of hydrodynamic type, our conjecture is a well-known result: any diagonalizable semi-Hamiltonian system possesses infinitely many conservation laws and commuting flows of hydrodynamic type, and can be solved by the generalized hodograph method \cite{Tsarev}.

\section{Classification of diagonalizable chains of the type (\ref{max})}

The structure of  equations (\ref{max}) implies that the only nonzero components $H^i_{jk}$ of the Haantjes tensor  are the ones with  $j\leq i+4, \ k\leq i+4$. Taking into account the skew-symmetry of the Haantjes tensor in its lower indices, this leaves ten essentially different components  of the type $H^1_{jk}$. Equating them to zero we obtain the expressions for {\it all} of the ten second order partial derivatives of $h(u^1, u^2, u^3, u^4)$:

\begin{eqnarray}
&&h_{11}=\frac{2g_1g_{12}-g_2g_{11}+2h_1g_{13}}{g_3}, \notag \\
&&h_{12}=\frac{g_{22}g_1+g_{11}+g_{13}h_2+g_{23}h_1}{g_3},\notag \\
&&h_{22}=\frac{g_2g_{22}+2g_{12}+2g_{23}h_2}{g_3} , \notag \\ 
&&h_{13} = \frac{g_{13}(h_3-g_2)+g_{23}g_1+g_{12}g_3+g_{33}h_1}{g_3}, \label{h} \\
&&h_{23}= g_{22}+\frac{g_{13}+h_3g_{23}+h_2g_{33}}{g_3},  \notag \\
&& h_{33} =2g_{23}-\frac{g_{33}(g_2-2h_3)}{g_3},  \notag \\ 
&&h_{14}= \frac{h_4g_{13}}{g_3}, ~~ h_{24}=\frac{h_4g_{23}}{g_3}, ~~ h_{34} = \frac{h_4g_{33}}{g_3},  ~~ h_{44} = 0. \notag 
\end{eqnarray}
Notice that these equations can be compactly written as
$$
d^2h=\frac{2}{g_3}(dhdg_3+dg_1du^2+dg_2dg-\frac{1}{2}g_2d^2g);
$$
here both sides of the equality are understood as symmetric two-forms, and $dg$, $dh$, $d^2g$, $d^2h$ denote the first and second symmetric differentials of $g$ and $h$.
The  consistency conditions for the equations (\ref{h}) lead to closed-form  expressions for {\it all} third order partial derivatives of the function $g(u^1, u^2, u^3)$ in terms of its lower order derivatives:
\begin{eqnarray}
&&g_{333}=\frac{2g_{33}^{2}}{g_{3}},\quad
g_{133}=\frac{2g_{13}g_{33}}{g_{3}}, \quad
g_{233}=\frac{2g_{23}g_{33}}{g_{3}},  \notag \\
&&g_{113}=\frac{2g_{13}^{2}}{g_{3}}, \quad
g_{123}=\frac{2g_{13}g_{23}}{g_{3}}, \quad
g_{223}=\frac{2g_{23}^{2}}{g_{3}},  \notag \\
&&g_{222}=\frac{2}{g_{3}^{2}}\left(
g_{2}g_{23}^{2}+g_{23}(g_{3}g_{22}+2g_{13})-g_{33}(g_{2}g_{22}+2g_{12})
\right),  \notag \\
&&g_{122}=\frac{2}{g_{3}^{2}}\left(
g_{1}g_{23}^{2}+g_{13}(g_{3}g_{22}+g_{13})-g_{33}(g_{1}g_{22}+g_{11})\right),
\label{g} \\
&&g_{112}=\frac{2}{g_{3}^{2}}\left(
g_{33}(g_{2}g_{11}-2g_{1}g_{12})-g_{13}(g_{2}g_{13}-2g_{3}g_{12})-g_{23}(g_{3}g_{11}-2g_{1}g_{13})\right),
\notag \\
&&g_{111}=\frac{2}{g_{3}^{2}}\left(
(g_{1}+g_{2}^{2})g_{13}^{2}+g_{1}^{2}g_{23}^{2}+g_{3}^{2}(g_{12}^{2}-g_{11}g_{22})-g_{22}g_{33}g_1^2 \right.
\notag \\
&&\qquad
+g_{13}g_{3}(g_{11}+2(g_{1}g_{22}-g_{2}g_{12}))+2g_{23}(g_{2}(g_{3}g_{11}-g_{1}g_{13})-g_{1}g_{3}g_{12})
\notag \\
&&\qquad -\left.
g_{33}((g_{1}+g_{2}^{2})g_{11}-2g_{1}g_{2}g_{12})\right). \notag
\end{eqnarray}
This system is in involution and its general solution depends on ten
integration constants, indeed, the values of $g$ and its partial derivatives
up to the second order can be prescribed arbitrarily at any fixed point 
$u^1_{0}, u^2_{0}, u^3_{0}$. The system (\ref{g}) was first derived in \cite{Pavlov2} from the requirement that the chain (\ref{max}) is embedded into a hierarchy of commuting hydrodynamic chains of  Egorov's type. Exactly the same equations for $g$ were obtained in \cite{Fer7} by applying the method of hydrodynamic reductions to the (2+1)-dimensional PDE
\begin{equation}
u_{tt}=g(u_{xx}, u_{xt}, u_{xy})
\label{eqmax}
\end{equation} 
which is naturally associated with the chain (\ref{max}); here the function $g$  is the same as in (\ref{max}), (\ref{g}). Thus, for hydrodynamic chains of the type (\ref{max}) the condition of diagonalizability is necessary and sufficient for the integrability. 

One can show that the vanishing of other components of the Haantjes tensor does not impose any additional constraints on the derivatives of $g$ and $h$. Thus, writing the fourth equation of the chain (\ref{max}) in the form $u^4_t=s(u^1, u^2, u^3, u^4, u^5)_x$ and setting $H^2_{jk}=0$, one obtains the expressions for {\it all} second order partial derivatives of $s$ in terms of $h$ and $g$, which are analogous to (\ref{h}). The consistency conditions   are satisfied identically modulo (\ref{h}), (\ref{g}).
Similarly, the condition $H^3_{jk}=0$ specifies the right hand side of the fifth equation of the chain, etc. 
Although we know no direct way to demonstrate the non-obstructedness of this recursive procedure in general, there exists an alternative direct approach to the reconstruction of a chain from the function $g(u^1, u^2, u^3)$. To illustrate this procedure we consider a simple example $g=u^3-\frac{1}{2}(u^1)^2$ which automatically satisfies (\ref{g}). The corresponding $h$, as specified by (\ref{h}),  is given by $h=\mu + \alpha u^1+\beta u^2+\gamma u^3+\delta u^4-u^1u^2$, which can be reduced to a canonical form $h= u^4-u^1u^2$ by redefining $u^4$ appropriately (this transformation  freedom  allows one  to absorb  arbitrary integration constants arising at each step of the construction). Thus, the first three equations of the chain are
\begin{equation}
u^1_t=u^2_x, ~~~ u^2_t=(u^3-\frac{1}{2}(u^1)^2)_x, ~~~ u^3_t=(u^4-u^1u^2)_x, ...,
\label{Benneycons}
\end{equation}
etc. Equations (\ref{Benneycons})  are nothing but the first three equations in the conservative representation of the Benney chain (\ref{Benney}); notice that the variables $u^i$ in (\ref{Benney}) and (\ref{Benneycons}) are not the same: $u^i$ in (\ref{Benneycons}) are conserved quantities of the chain (\ref{Benney}). The recostruction of the remaining equations of the chain consists of three steps.

\noindent (i) One introduces the corresponding PDE (\ref{eqmax}): in our case this will be a potential form of the dispersionless KP equation, $u_{tt}=u_{xy}-\frac{1}{2}u_{xx}^2$. It was demonstrated in \cite{Fer7} that the general PDE (\ref{eqmax}) is integrable by the method of hydrodynamic reductions if and only if the function $g$ satisfies the relations (\ref{g}). 

\noindent (ii) One constructs a dispersionless Lax pair for the PDE (\ref{eqmax}), in our example it takes the form
$$
p_t=(\frac{1}{2}p^2+u_{xx})_x, ~~~ p_y=(\frac{1}{3}p^3+u_{xx}p+u_{xt})_x;
$$
the consistency conditions $p_{ty}=p_{yt}$ are satisfied identically modulo the dispersionless KP equation. The existence of such Lax pairs  was established in \cite{Pavlov2, Fer7} for  any equation (\ref{eqmax}) provided $g$ satisfies the compatibility conditions (\ref{g}). 

\noindent (iii) One looks for $p$ as an expansion in the auxiliary parameter $\lambda$,
$$
p=\lambda-\frac{u^1}{\lambda}-\frac{u^2}{\lambda^2}-\frac{u^3}{\lambda^3}-....;
$$
the substitution of this ansatz into the first equation of the Lax pair implies an infinite hydrodynamic chain for the variables $u^i$. The first three equations of this chain identically coincide with (\ref{Benneycons}). The substitution into the second equation of the Lax pair produces a commuting chain (one has to set $u^1=u_{xx}, \ u^2=u_{xt}$). 
Both chains  possess infinitely many hydrodynamic reductions since this is the case for the generating equation (\ref{eqmax}). Thus, the Haantjes tensor will automatically vanish (as demonstrated in Sect. 5). 

This procedure has been successfully implemented in \cite{Pavlov6} for various classes of solutions of the system (\ref{g}), namely, cases (\ref{g1}) and (\ref{g2}) of the classification presented below (according to \cite{Pavlov2}, the case (\ref{g2}) is reciprocally related to the case (\ref{g1}) and, therefore, does not require a special treatment). The work on (\ref{g3}) and (\ref{g4}) is currently in progress \cite{Pavlov4}.

\subsection{Integration of the system (\ref{g})}

To explicitly calculate $g(u^1, u^2, u^3)$  we will follow \cite{Pavlov2}. The main  observation is that the first six equations in (\ref{g}) imply
that the function $1/g_3$ is linear,
$1/g_3=\alpha+\beta u^1+\gamma u^2+\delta u^3$. If $\delta \ne 0$ then, up to a linear change of variables, one can assume  that $1/g_3=u^3$. Similarly, if $\delta=0, \ \gamma \ne 0$, one can set  $1/g_3=u^2$. If $\delta=\gamma=0, \ \beta \ne 0$ one has $1/g_3=u^1$. The last possibility is $1/g_3=1$. Thus, we have four   cases to consider:
$$
g=u^3+p(u^1, u^2), ~~~  g=\frac{ u^3}{u^1}+p(u^1, u^2), ~~~ g=\frac{u^3}{u^2}+p(u^1, u^2), ~~~ g=\ln u^3+p(u^1, u^2);
$$
here the function $p(u^1, u^2)$ can be recovered  after the substitution  into the remaining four equations (\ref{g}). In each of these cases the resulting equations for $p(u^1, u^2)$ integrate explicitly, see \cite{Pavlov2}, leading to the four essentially different canonical forms:
\begin{eqnarray}
g &=&u^3+\frac{1}{4A}(Au^2+2Bu^1)^{2}+Ce^{-Au^1},  \label{g1} \\
g &=&\frac{u^3}{u^1}+\left( \frac{1}{u^1}%
-\frac{A}{4(u^1)^{2}}\right) (u^2)^{2}+\frac{B}{(u^1)^{2}}u^2-\frac{B^{2}}{A(u^1)^{2}}-Ce^{-A/u^1},
 \label{g2} \\
g &=&\frac{u^3}{u^2}+\frac{1}{6}\eta (u^1)(u^2)^{2},  \label {g3} \\
g &=&\ln u^3-\ln \sigma \left(u^1, u^2\right) -\frac{1}{4}{\int }\eta (u^1)du^1. \label{g4}
\end{eqnarray}%
Here  $\eta (u^1)$ is a solution to the Chazy equation \cite{Chazy},
\begin{equation}
\eta
^{\prime \prime \prime }+2\eta \eta ^{\prime \prime }-3\eta
^{\prime ^{2}}=0,
\label{Chazy}
\end{equation}
and $\sigma (u^1, u^2)$ is an elliptic sigma function in the variable $u^2$ whose dependence on $u^1$ is governed by  the Chazy equation (see Case IV below).
Details of the derivation of canonical forms (\ref{g1}) - (\ref{g4}) can be summarized as follows.

\noindent {\bf Case I}. Substituting the ansatz $g=u^3+p(u^1, u^2)$ into (\ref{g}) one arrives at the  equations 
\begin{eqnarray}
&&p_{111}=2(p_{12}^2-p_{11}p_{22}), \notag \\
&&p_{112}=p_{122}=p_{222}=0. \notag
\end{eqnarray}
The last three equations imply 
$$
p=\frac{1}{4}A(u^2)^2+(Bu^1+D)u^2+q(u^1),
$$
 and the substitution into the first equation results in the linear ODE
$q'''+Aq''=2B^2$. Up to a transformation of the form  $u^3 \to u^3+\alpha u^2+\beta u^1+ \gamma$  this leads to (\ref{g1}).

\noindent {\bf Case II}. Substituting the ansatz $g=u^3/u^1+p(u^1, u^2)$ into (\ref{g}) one arrives at the  equations 
\begin{eqnarray}
&&p_{111}=2(p_{12}^2-p_{11}p_{22})+\frac{2}{(u^1)^2}(p_1+p_2^2)-
\frac{2}{u^1}(p_{11}+2p_1p_{22}-2p_2p_{12}), \notag \\
&&p_{112}=-\frac{2}{(u^1)^2}p_2-
\frac{4}{u^1}p_{12}, \notag \\
&&p_{122}=\frac{2}{(u^1)^2}-
\frac{2}{u^1}p_{22},  ~~~ p_{222}=0.\notag
\end{eqnarray}
The last three equations imply 
$$
p=\left(\frac{1}{u^1}-\frac{A}{4(u^1)^2}\right)(u^2)^2+\left(\frac{D}{u^1}+\frac{B}{(u^1)^2}\right)u^2
-\frac{B^2}{A(u^1)^2}+q(u^1),
$$
and the substitution into the first equation results in the linear ODE
$(u^1)^3q'''+u^1(6u^1-A)q''+(6u^1-2A)q'=0$ whose  basis of  solutions consists of $1, \ 1/u^1$ and $e^ {-A/u^1}$. Up to a transformation of the form $u^3 \to u^3+\alpha u^2+\beta u^1+ \gamma$ this implies (\ref{g2}). It was observed in \cite{Pavlov2} that the cases I and  II are reciprocally related: under the change from $x, t$ to the new independent variables $X,  T $ defined as
$dX=u^1dx+u^2dt, \ T=t$, and the introduction of  the new dependent variables $U^1=\frac{1}{u^1}, \ U^2=-\frac{u^2}{u^1}, \ U^3=-\frac{u^3}{u^1},$  etc, the chains from the case I transform to the chains from the case II, and vice versa. On the level of the corresponding equations  this means that the change of variables 
$U^1=\frac{1}{u^1}, \ U^2=-\frac{u^2}{u^1}, \ P=\frac{(u^2)^2}{u^1}-p, \ G=\frac{(u^2)^2}{u^1}-g$ transforms the equations for $p$ from the Case I to the equations for $p$ from the Case II. Equivalently, 
(\ref{g1}) goes to (\ref{g2}).

\noindent {\bf Case III}. Substituting the ansatz $g=u^3/u^2+p(u^1, u^2)$ into (\ref{g}) one arrives at the  equations 
\begin{eqnarray}
&&p_{111}=\frac{2}{(u^2)^2}p_1^2+2(p_{12}^2-p_{11}p_{22})+\frac{4}{u^2}(p_1p_{12}-p_2p_{11}),
\notag \\
&&p_{112}=\frac{2}{u^2}p_{11}, ~~~
p_{122}=\frac{2}{(u^2)^2}p_{1},  \notag \\
&& p_{222}=\frac{2}{(u^2)^2}p_{2}-\frac{2}{u^2}p_{22}. \notag
\end{eqnarray}
The last three equations imply 
$$
p=A+\frac{B+Cu^1}{u^2}+\frac{1}{6}\eta (u^1)(u^2)^2,
$$
and the substitution into the first relation results in the Chazy equation
(\ref{Chazy}) for $\eta$. Elimination of the constants $A, B, C$ leads to (\ref{g3}).

\noindent {\bf Case IV}. Substituting the ansatz $g=\ln u^3+p(u^1, u^2)$ into (\ref{g}) one arrives at the  equations 
\begin{eqnarray}
&&p_{111}=2(p_{12}^2-p_{11}p_{22})+2(p_1+p_2^2)p_{11}-4p_1p_2p_{12}+2p_1^2p_{22},
\notag \\
&&p_{112}=4p_1p_{12}-2p_2p_{11}, \notag \\
&&p_{122}=2p_{11}+2p_1p_{22},  \notag \\
&& p_{222}=4p_{12}+2p_2p_{22}. \notag
\end{eqnarray}
The general solution of the fourth equation can be represented in the form
$$
p=-\ln \sigma \left(u^1, u^2\right) -\frac{1}{4}{\int }\eta (u^1)du^1;
$$
here $\sigma$ solves the heat equation $4\sigma_1=\sigma_{22}$ and $\eta$ is a function of $u^1$. 
It is convenient to introduce the new variable $v(u^1, u^2)$ by the formula $v=-(\ln \sigma)_{22}$. Taking into account the heat equation for $\sigma$ one has
$$
v_1=\frac{1}{4}v_{22}-\frac{1}{2}v^2+\frac{1}{2}(\ln \sigma)_2v_2.
$$
Rewritten in terms of $v$, the third equation for $p$ implies 
\begin{equation}
v_{22}=6v^2-4v\eta-4\eta',
\label{v22}
\end{equation}
$'\equiv d/du^1$, the second equation is satisfied identically and the first takes the form
\begin{equation}
v_2^2=4v^3-4v^2\eta-8v\eta'-\frac{8}{3}\eta''.
\label{v2}
\end{equation}
This shows that $v$ is a shift of the Weierstrass elliptic function in the variable $u^2$.  Since $v=-(\ln \sigma)_{22}$, the function $\sigma$ is the corresponding  theta function. Notice that (\ref{v22}) can be obtained as a result of differentiation of (\ref{v2}) by  $u^2$. Thus, we have two equations for $v$:
\begin{eqnarray}
&&v_1=v^2-v\eta-\eta'+\frac{1}{2}(\ln \sigma)_2v_2, \notag \\
&&v_2^2=4v^3-4v^2\eta-8v\eta'-\frac{8}{3}\eta''. \notag 
\end{eqnarray}
The condition of their consistency leads to the Chazy equation for $\eta$.

\section{Classification of diagonalizable chains of the type (\ref{max1})}

In this section we sketch the classification of diagonalizable chains of the type (\ref{max1}). The condition $H^1_{jk}=0$ implies the following formulae for  second order partial derivatives of $h$:
\begin{eqnarray}
&&h_{14}= \frac{h_4g_{13}}{g_3} , ~~~ h_{24}=\frac{h_4g_{23}}{g_3} , ~~~ h_{34} = \frac{h_4g_{33}}{g_3},  ~~~ h_{44} = 0, \notag \\
&&h_{13} =-\frac{f_{22}g_1}{f_2}+ \frac{g_{13}(h_3-g_2)+g_{23}g_1+g_{12}g_3+g_{33}h_1}{g_3}, \notag \\
&&h_{23}= g_{22}-f_{12}+\frac{(f_1-g_2)f_{22}}{f_2}+\frac{(h_3-f_1)g_{23}+f_2g_{13}+h_2g_{33}}{g_3}, \notag \\
&&h_{33} =2g_{23}-\frac{f_{22}g_3}{f_2}-\frac{g_{33}(f_1+g_2-2h_3)}{g_3}, \label{hh} \\ 
&&h_{22}=-\frac{f_{22}(f_1^2-2f_1g_2+g_2^2+f_2g_1)}{g_3f_2} +\frac{(g_2-f_1)(g_{22}-2f_{12})+f_2(2g_{12}-f_{11})+2g_{23}h_2}{g_3}, \notag \\
&&h_{12}=\frac{f_{22}g_1(f_1-g_2)}{g_3f_2}+\frac{g_{22}g_1+f_2g_{11}+g_{13}h_2+g_{23}h_1-2g_1f_{12}}{g_3}, \notag \\
&&h_{11}=-\frac{g_1^2f_{22}}{g_3f_2}+\frac{2g_1g_{12}+(f_1-g_2)g_{11}+2h_1g_{13}-g_1f_{11}}{g_3}. \notag
\end{eqnarray}
By calculating the consistency conditions for the above equations we  obtain the expressions for all third order partial derivatives of $g$ and $f$.

\noindent {\bf Equations for $g$:}
\begin{eqnarray}
&&g_{333}=\frac{2g_{33}^{2}}{g_{3}},\quad
g_{133}=\frac{2g_{13}g_{33}}{g_{3}}, \quad
g_{233}=\frac{2g_{23}g_{33}}{g_{3}},  \notag \\
&&g_{113}=\frac{2g_{13}^{2}}{g_{3}}, \quad
g_{123}=\frac{2g_{13}g_{23}}{g_{3}}, \quad
g_{223}=\frac{2g_{23}^{2}}{g_{3}},  \notag \\
\ \notag \\
&&g_{111} =\frac{f_{11} (-g_{33}g_2g_1+g_3(g_{23}g_1+g_2g_{13}-g_3g_{12}))
+2f_1^2(g_{13}^2-g_{33}g_{11})+f_1f_{11}(g_{33}g_1-g_3g_{13})}{f_2g_3^2}\notag\\
&&+2\frac{2f_1((-g_{33}g_1+g_3g_{13})g_{12}+g_{23}(g_1g_{13}-g_3g_{11})+g_2(-g_{13}^2+g_{33}g_{11}))+g_{23}^2g_1^2}{f_2g_3^2} \notag \\
&&+2\frac{g_{22}(-g_1^2g_{33}+2g_3g_1g_{13}-g_3^2g_{11})
+g_{13}^2(g_2^2+f_2g_1)+g_{12}(2g_{33}g_2g_1-2g_2g_3g_{13}+g_3^2g_{12})}{f_2g_3^2} \notag\\
&&+2\frac{g_{11}(-g_{33}g_2^2-f_2g_{33}g_1+f_2g_3g_{13})
-2g_{23}(g_1g_3g_{12}+g_2(g_1g_{13}-g_3g_{11}))}{f_2g_3^2} \notag\\
&&+2\frac{f_{12}(g_{33}g_1^2 +g_3(-2g_1g_{13}+g_3g_{11}))}{f_2g_3^2}, \label{gf1} \\
&&g_{112} = \frac{(g_{33}g_1^2+g_3(g_3g_{11}-2g_1g_{13}))f_{22}
+f_2(f_{11}(g_{33}g_1-g_3g_{13})-2g_{13}(g_2g_{13}-2g_3g_{12}))}{f_2g_3^2}\notag\\
&&-2\frac{g_{33}(2g_1g_{12}-g_2g_{11})
+g_{23}(g_3g_{11}-2g_1g_{13})+f_1(g_{33}g_{11}-g_{13}^2)}{g_3^2},\notag
\end{eqnarray}
\begin{eqnarray}
&&g_{122} = f_{22}\frac{(g_1g_2g_{33}+(g_3g_{13}-g_1g_{33})f_1+(g_3g_{12}-g_{23}g_1-g_2g_{13})g_3)
}{f_2g_3^2}\notag\\
&&+\frac{2(g_{23}^2g_1-g_{33}g_{22}g_1+g_3g_{22}g_{13}
+f_2g_{13}^2+(g_{33}g_1-g_3g_{13})f_{12}-f_2g_{33}g_{11})}{g_3^2},\notag\\
\ \notag \\
&&g_{222} = \frac{(((f_1-2g_2)g_{33} +2g_3g_{23})f_1 +(g_2^2+f_2g_1)g_{33}+(g_3g_{22}-2g_2g_{23}-f_2g_{13})g_3)f_{22}}{f_2g_3^2}\notag\\
&&+2\frac{-f_1(f_{12}g_{33}+g_{23}^2-g_{33}g_{22})+f_{12}(g_{33}g_2-g_3g_{23})+g_3g_{23}g_{22}
+g_2(g_{23}^2-g_{22}g_{33})}{g_3^2} \notag\\
&&+\frac{f_2(g_{33}(f_{11}-4g_{12})+4g_{23}g_{13})}{g_3^2}. \notag 
\end{eqnarray}
\noindent {\bf Equations for $f$:}
\begin{eqnarray}
&&f_{111} = -\frac{f_{11}(g_{33}(f_1^2+g_2^2+f_2g_1)-2g_2g_3g_{23}+g_3^2g_{22}-f_2g_3g_{13})
-2f_1f_{11}(g_{33}g_2-g_3g_{23}) }{f_2g_3^2}\notag\\
&&+f_{12}\frac{g_3^2f_{11}+2(g_1g_2g_{33}+f_1(-g_{33}g_1+g_3g_{13})
+g_3(-g_{23}g_1-g_2g_{13}+g_3g_{12}))}{f_2g_3^2}\notag\\
&&-\frac{f_{22}(g_{33}g_1^2-2g_1g_3g_{13}+g_3^2g_{11})}{f_2g_3^2},\label{gf2} \\
\ \notag \\
&&f_{112} = \frac{f_{22}f_{11}g_3^2 - f_2 (f_1f_{11}g_{33}+f_{11}(g_3g_{23}-g_{33}g_2) +2f_{12}(g_{33}g_1-g_3g_{13}))}{f_2g_3^2},\notag\\
\ \notag \\
&&f_{122} = \frac{-f_2^2f_{11}g_{33}+f_{22}(f_{12}g_3^2+f_2(-g_{33}g_1+g_3g_{13}))}{f_2g_3^2},\notag\\
\ \notag \\
&&f_{222} = \frac{f_{22}^2g_3^2-2f_2^2f_{12}g_{33}+f_2f_{22}((f_1-g_2)g_{33}+g_3g_{23})}{f_2g_3^2}.\notag
\end{eqnarray}
We have verified that the system (\ref{gf1}), (\ref{gf2}) is in involution. We claim that the functions $f$ and $g$  contain all the necessary information about conservative chains (\ref{max1}). In particular, equations (\ref{hh}) allow one to reconstruct the function $h$ and, hence, the right hand side of the third equation of the chain. Similarly, the requirement $H^2_{jk}=0$ reconstructs the fourth equation (i.e., provides an involutive second order system for  the right hand side of the fourth equation), etc. Although we have verified directly that our procedure is non-obstructed up to order 5, we could not establish this property in general. Possible approaches to this problem could be

\noindent --- developing a cohomological approach to the  reconstruction procedure in the spirit of \cite{DubZ};

\noindent --- establishing a Hamiltonian formulation for general integrable chains of the type (\ref{max1});

\noindent --- establishing a link of integrable chains of the type (\ref{max1}) to $(2+1)$-dimensional integrable PDEs, as explained in Sect. 2. Although there is a general belief that one can construct a one-to-one correspondence between integrable hydrodynamic chains and integrable $(2+1)$-dimensional quasi-linear PDEs (that is, any integrable PDE can be `decoupled' into a pair of commuting chains), there exists yet no rigorous proof of this statement.

\subsection{Integration of the equations (\ref{gf1}), (\ref{gf2})}

Notice that the first six equations in (\ref{gf1}) are exactly the same as in Sect. 2.
Thus, there are four essentially different cases to consider.

\medskip

\noindent {\bf Case I:} $g=u^3+p(u^1, u^2)$.  Substituting this ansatz  into (\ref{gf1}), (\ref{gf2})  one arrives at the  following relations.

\noindent {\bf equations for $p$:} 
\begin{eqnarray*}
&&p_{111}= \frac{p_{12}(2p_{12}-f_{11})+2p_{11}(f_{12}-p_{22})}{f_2},\\
&&p_{112}=\frac{f_{22}p_{11}}{f_2}, \quad p_{122}=\frac{f_{22}p_{12}}{f_2},   \quad p_{222} = \frac{f_{22}p_{22}}{f_2};
\end{eqnarray*}

\noindent {\bf equations for $f$:} 
\begin{eqnarray*}
&&f_{111}=\frac{f_{11}(f_{12}-p_{22})+2f_{12}p_{12}-f_{22}p_{11}}{f_2}, \\
&&f_{112}=\frac{f_{22}f_{11}}{f_2},\quad f_{122}=\frac{f_{22}f_{12}}{f_2}, \quad f_{222}=\frac{f_{22}^2}{f_2}.
\end{eqnarray*}
The last three equations for $f$ and the last three equations for $p$ lead, up to elementary changes of variables, to the two  possibilities.

\noindent {\bf Subcase ${\rm I}_1$:} 
$$
f=s(u^1)e^{u^2}, ~~~ p=q(u^1)e^{u^2}.
$$
The substitution of these expressions into the  remaining equations for $p_{111}$ and $f_{111}$ leads to a system of coupled ODEs for $s(u^1)$ and $q(u^1)$:
\begin{eqnarray*}
&&q'''=\frac{2((q')^2-qq'')-q's''+2q''s'}{s}, \\
&&s'''= \frac{s''s'+2s'q'-q s''-q'' s}{s}.
\end{eqnarray*}
Setting  $q=-s'$, the second equation will be satisfied identically while  the first  implies a fourth order ODE 
 $s''''s+3(s'')^2-4s's'''=0$ whose general solution is an elliptic sigma-function: $s=\sigma (u^1)$, here $(\ln \sigma)''=-\wp, \  (\wp')^2=4\wp^3-c$
(notice that $g_2=0, \ g_3=c$). Thus,  as a particular case we have
$$
f=\sigma(u^1)e^{u^2}, ~~~ p=-\sigma '(u^1)e^{u^2}.
$$

\noindent {\bf Subcase ${\rm I}_{2}$:} 
$$
f=(a u^1+b)u^2 +s(u^1), ~~~ p=\frac{1}{2}A (u^2)^2+B u^1u^2+q(u^1),
$$
$a, b, A, B$=const. The substitution of these expressions into the  remaining equations for $p_{111}$ and $f_{111}$ leads to linear ODEs for $s(u^1)$ and $q(u^1)$:
\begin{eqnarray*}
&&q'''=\frac{2B^2-B s''+2(a-A)q''}{au^1+b}, \\
&&s'''=\frac{(a-A)s''+2aB}{au^1+b}.
\end{eqnarray*}
These equations are straightforward to solve. One needs to consider two different cases: $a=0, \ b=1$ and $a=1, \ b=0$. If $a=0, \ b=1$ then, up to unessential integration constants, we have
$$
s=\alpha e^{-Au^1}, ~~~ q=\frac{B^2}{2A}(u^1)^2-\alpha \frac{B}{A}e^{-Au^1}+\beta e^{-2Au^1}.
$$
The case $a=1, \ b=0$ leads to
$$
s=\frac{B}{A-1}(u^1)^2+\alpha (u^1)^{3-A}, ~~~
q=\frac{B^2(A-2)}{2(A-1)^2}(u^1)^2+\frac{\alpha B}{1-A}(u^1)^{3-A}+\beta (u^1)^{2(2-A)}.
$$

\medskip

\noindent {\bf Case II:} $g=u^3/u^1+p(u^1, u^2)$. Substituting this ansatz  into (\ref{gf1}), (\ref{gf2})  one arrives at the  following set of relations.

\noindent {\bf equations for $p$:} 
\begin{eqnarray*}
&&p_{111}= 2\frac{p_{2}^2+f_{1}^2+f_2p_1-4p_2f_1}{f_2(u^1)^2} +\frac{(f_1-p_2)(f_{11}-4p_{12})+4p_1(f_{12}-p_{22})}{u^1f_2}+\\
&&\frac{p_{12}(2p_{12}-f_{11})+2p_{11}(f_{12}-p_{22})}{f_2}-2\frac{p_{11}}{u^1},\\
&&p_{112}=\frac{2}{(u^1)^2}(f_1-p_2)   +\frac{f_{11}-4p_{12}}{u^1} +\frac{2p_1f_{22}}{u^1f_2}
+\frac{f_{22}p_{11}}{f_2}, \\
&&p_{122}= 2\frac{f_2}{(u^1)^2}+\frac{2}{u^1}(f_{12}-p_{22})+\frac{f_{22}(p_2-f_1)}{u^1f_2}+\frac{f_{22}p_{12}}{f_2},\\
&&p_{222} = \frac{f_{22}p_{22}}{f_2} +\frac{f_{22}}{u^1};
\end{eqnarray*}

\noindent {\bf equations for $f$:}
\begin{eqnarray*}
&&f_{111}=\frac{f_{11}}{f_2}(f_{12}-p_{22}-\frac{f_2}{u^1}) +\frac{f_{12}}{f_2u^1}(2p_2-2f_1+p_{12}u^1)-\frac{f_{22}}{f_2u^1}(2p_1+p_{11}u^1), \\
&&f_{112}=\frac{f_{11}f_{22}}{f_2}-2\frac{f_{12}}{u^1},\quad
f_{122}=\frac{f_{12}f_{22}}{f_2}-\frac{f_{22}}{u^1},\quad
f_{222}=\frac{f_{22}^2}{f_2}.
\end{eqnarray*}
The cases I and  II are reciprocally related (we thank  Maxim Pavlov for pointing out  this equivalence): under the change from $x, t$ to the new independent variables $X,  T $ defined as
$dX=u^1dx+f dt, \ T=t,$ and the introduction of  the new dependent variables $U^1=\frac{1}{u^1}, \ U^2=-\frac{u^2}{u^1}, \ U^3=-\frac{u^3}{u^1}, \ $  etc, the chains from the case I transform to the chains from the case II, and vice versa. On the level of the corresponding equations  this means that the change of variables 
$U^1=\frac{1}{u^1}, \ U^2=-\frac{u^2}{u^1}, \ F=-\frac{f}{u^1}, \ P=\frac{u^2}{u^1}f -p$ transforms the equations for $p, f$ from the Case I to the equations for $P, F$ from the Case II. Thus, we will not discuss this case in any more detail here. 

\medskip

\noindent {\bf Case III:} $g=u^3/u^2+p(u^1, u^2)$.  Substituting this ansatz  into (\ref{gf1}), (\ref{gf2})  one arrives at the  following set of relations.

\noindent {\bf equations for $p$:} 
\begin{eqnarray*}
&&p_{111}=\frac{2p_1^2}{f_2(u^2)^2}+\frac{p_1(4p_{12}-f_{11})+4p_{11}(f_1-p_2)}{u^2f_2}+\frac{p_{12}(2p_{12}-f_{11})+2p_{11}(f_{12}-p_{22})}{f_2},\\
&&p_{112}=\frac{f_{22}p_{11}}{f_2}+2\frac{p_{11}}{u^2},\quad
p_{122}=\frac{2f_2p_1+u^2f_{22}(p_1+u^2p_{12})}{f_2(u^2)^2},\\
&&p_{222}=2\frac{f_{22}(p_2-f_1)-f_2p_{22}}{u^2f_2}+\frac{p_{22}f_{22}}{f_2}+2\frac{f_{12}}{u^2}-\frac{2}{(u^2)^2}(f_1-p_2);
\end{eqnarray*}

\noindent  {\bf equations for $f$:} 
\begin{eqnarray*}
&&f_{111}= \frac{f_{11}}{f_2u^2}(2(f_1-p_2)-p_{22}u^2) +\frac{f_{12}}{f_2u^2}(u^2(f_{11}+2p_{12})+2p_1)-\frac{f_{22}p_{11}}{f_2},\\
&&f_{112}=\frac{f_{22}f_{11}}{f_2} +\frac{f_{11}}{u^2},\quad
f_{122}=\frac{f_{22}f_{12}}{f_2},\quad
f_{222}=\frac{f_{22}^2}{f_2}-\frac{f_{22}}{u^2}.
\end{eqnarray*}
The last three equations for $f$ and the last three equations for $p$ lead to the two essentially different possibilities:

\noindent {\bf Subcase ${\rm III}_1$:} 
$$
f=s(u^1)(u^2)^k, ~~~ p=q(u^1)(u^2)^{k+1},
$$
$k$=const. The substitution of these expressions into the  remaining equations for $p_{111}$ and $f_{111}$ leads to the coupled system of ODEs for $s(u^1)$ and $q(u^1)$:
\begin{eqnarray*}
&&q'''=\frac{k+2}{ks}(2(k+2)(q')^2-2(k+1)q q''-q's''+2q''s'),\\
&&s'''=\frac{k+2}{ks}(s''s'-k(k+1)s''q+2ks'q')-(k-1)q''.
\end{eqnarray*}
Notice that under the substitution $s=k=1$ the equation for $q$ reduces to the Chazy equation (\ref{Chazy}) for $q(u^1)=\frac{1}{6}\eta(u^1)$. 

\noindent {\bf Subcase ${\rm III}_{2}$:} 
$$
f=(au^1+b)\ln u^2 +s(u^1), ~~~ p=q (u^1)u^2+a(u^2\ln u^2-u^2),
$$
$a, b$=const. The substitution of these expressions into the  remaining equations for $p_{111}$ and $f_{111}$ leads to the coupled system of ODEs for $s(u^1)$ and $q(u^1)$:
\begin{eqnarray*}
&&q'''=\frac{8(q')^2-4q q''-2q's''+4q''s'}{au^1+b}, \\
&&s'''=2\frac{s''(s'-q)+2aq'}{au^1+b}+q''.
\end{eqnarray*}

\medskip

\noindent {\bf Case IV:} $g=\ln u^3+p(u^1, u^2)$.  Substituting this ansatz  into (\ref{gf1}), (\ref{gf2}) one obtains

\noindent {\bf equations for $p$:} 
\begin{eqnarray*}
&&p_{111}=\frac{p_{22}(2p_1^2-2p_{11})+p_{11}(p_2(2p_2-4f_1)+2(f_1^2+f_{12}+f_2p_1))+p_{12}(2p_{12}-f_{11}+4p_1(f_1-p_2))}{f_2}\\
&&-\frac{p_1(2p_1f_{12}+(f_1-p_2)f_{11})}{f_2},\\
&&p_{112}=\frac{f_{22}(p_{11}-p_1^2)}{f_2}+p_1(4p_{12}-f_{11})+2p_{11}(f_1-p_2),\\
&&p_{122}=\frac{f_{22}}{f_2}(p_1(f_1-p_2)+p_{12})+2p_1(p_{22}-f_{12})+2f_2p_{11},\\
&&p_{222}=\frac{f_{22}}{f_2}(p_{22}+f_1(2p_2-f_1)-p_2^2)+2(p_{22}-f_{12})(p_2-f_1)-f_{22}p_1+f_2(4p_{12}-f_{11});
\end{eqnarray*}

\noindent {\bf equations for $f$:} 
\begin{eqnarray*}
&&f_{111}= \frac{f_{11}}{f_2}(f_1^2+p_2^2+f_2p_1-p_{22}-2f_1p_2+f_{12}) +2\frac{f_{12}}{f_2}(p_1(f_1-p_2)+p_{12})+\frac{f_{22}}{f_2}(p_1^2-p_{11}),\\
&&f_{112}=\frac{f_{22}f_{11}}{f_2}+f_{11}(f_1-p_2)+2f_{12}p_1,\quad
f_{122}=\frac{f_{22}f_{12}}{f_2}+f_2f_{11}+f_{22}p_1,\\
&&f_{222}=\frac{f_{22}^2}{f_2}+2f_2f_{12}+f_{22}(p_2-f_1).
\end{eqnarray*}
Although this system  is in involution and reduces to the corresponding Case 4 in Sect. 2 under the substitution $f=u^2$, we were not able to  integrate it in general. Let us just mention that  the last three relations for $f$ imply the Monge-Ampere equation $f_{11}f_{22}-f_{12}^2=a(u^1)f_2^2$. This 
suggests  a separable ansatz  $f=s(u^1)r(u^2)$. A simple analysis leads to the two possibilities:

\noindent (i) $f=s(u^1)(u^2)^k, ~~ p=\frac{1}{1-k}s'(u^1)(u^2)^{k+1}-\ln u^2, $ where $s$ satisfies the equation $ss''-\frac{k}{k-1}(s')^2=0$. Without any loss of generality one can take $s=(u^1)^{1-k}$.

\noindent (ii) $f=s(u^1)e^{u^2}, ~~ p=-s'(u^1)e^{u^2}, $ where $s$ satisfies the equation $ss''-(s')^2=0$. Up to a linear transformation of $u^1$ one has $s=e^{u^1}$.

\section{Classification of diagonalizable Hamiltonain chains}

It was observed in \cite{Kup1, Kup2} that the Benney chain (\ref{Benney}) can be represented in the Hamiltonian form (\ref{Ham}),
$$
{\bf u}_t=\left (B\frac{d}{dx}+\frac{d}{dx}B^t \right )\frac{\partial h}{\partial {\bf u}}, 
$$
where $B^{ij}=(i-1)u_{i+j-2}$, and  $h=(u^3+(u^1)^2)/2$ is the Hamiltonian density. 
Further integrable examples can be constructed by looking at  Hamiltonian densities in the form $h=u^3+p(u^1, u^2)$, and imposing the constraint $H^1_{jk}=0$. This  implies  the relations 
$$
p_{111}(2+u^1p_{22})=p_{11}p_{22}-p_{12}^2, ~~~ p_{112}=p_{122}=p_{222}=0,
$$
which, up to a natural equivalence 
$h\to \alpha h +au^1+bu^2+c$,  lead to  the Hamiltonian densities 
\begin{equation}
h=u^3+\alpha (u^1)^2+\beta u^1u^2 +\gamma (u^2)^2+\delta (u^1)^3
\label{lin}
\end{equation}
where the constants  $\alpha, \beta, \gamma, \delta$ satisfy a single relation $\beta^2-4\alpha \gamma +12 \delta=0$. These densities were first introduced  in \cite{Kup3},  where it was shown that  the corresponding Hamiltonian chains possess an infinity of conservation laws which Poisson commute and form a complete system.  
One can prove that all other components $H^i_{jk}, \ i\geq 2$, of the Haantjes tensor are identically zero.

The aim of this section is to characterize {\it all}  densities of the form $h(u^1, u^2, u^3)$ such that the Haantjes tensor of the  Hamiltonian chain (\ref{Ham}) is zero. As in the previous Sections, the conditions $H^1_{jk}=0$ provide  expressions for all of the ten third order partial derivatives of $h$, the simplest six of them being
\begin{eqnarray}
&&h_{333}=\frac{5h_{33}^2}{2h_{3}},\quad
h_{133}=\frac{5h_{13}h_{33}}{2h_3},\quad
h_{233}=\frac{5h_{23}h_{33}}{2h_3},  \nonumber \\
&&h_{113}=\frac{3h_{13}^2+2h_{33}h_{11}}{2h_3}, \label{HH} \\
&&h_{123}=\frac{3h_{13}h_{23}+2h_{33}h_{12}}{2h_3}, \nonumber \\
&&h_{223}=\frac{3h_{23}^2+2h_{33}h_{22}}{2h_3};   \nonumber
\end{eqnarray}
the remaining expressions for  $h_{111}, h_{112}, h_{122}, h_{222}$ are not written out explicitly due to their complexity (we have used Mathematica to manipulate with these expressions). It was verified that the system for $h$ is in involution, and its solution space is 10-dimensional. In what follows, (\ref{HH}) refers to the full system of ten equations. It was demonstrated recently in \cite{GibRai} that the relations (\ref{HH}) imply the vanishing of all remaining components of the Haantjes tensor, that is,
$$
H^1_{jk}=0\implies H^i_{jk}=0 \ \ {\rm for\  any}\  i\geq 2.
$$
Before proceeding with a detailed analysis of the equations (\ref{HH}), let us make a digression on conservation laws of the chains (\ref{Ham}). First of all, any Hamiltonian chain (\ref{Ham}) possesses three conservation laws of the form
$$
u^1_t=(u^1h_2+2u^2h_3)_x,
$$
$$
u^2_t=(u^1h_1+2u^2h_2+3u^3h_3-h)_x,
$$ 
and
$$
h_t=(u^1h_1h_2+2u^2h_1h_3+u^2h_2^2+3u^3h_2h_3+2u^4h_3^2)_x,
$$
which correspond to the conservation of the Casimir, momentum and the Hamiltonian, respectively. Let us require the existence of an extra `higher' conservation law of the form
\begin{equation}
P(u^1, u^2, u^3, u^4)_t=Q(u^1, u^2, u^3, u^4, u^5)_x
\label{extra}
\end{equation}
whose density $P$ depends on the first four coordinates $u^i$ (the structure of the flux $Q$ follows from the equations of the chain and does not constitute an additional restriction). We have obtained the following

\medskip

\noindent {\bf Proposition.} {\it The relations (\ref{HH}) are necessary and sufficient for the existence of an extra conservation law of the form (\ref{extra}).}

\medskip

The proof is computational, see Appendix 2 for details. In fact, we have  the following stronger result

\begin{theorem} The relations (\ref{HH}) imply the existence of  a generating function which gives rise to an infinity of conservation laws. To be more precise,  we claim that for any $n$ there exist $n$ linearly independent conserved densities  which are functions of the first $n$ coordinates $u^i$. All higher conservation laws are polynomial in  $u^4, u^5, u^6, ...$. 
\end{theorem}

The proof of this statement is given in Sect. 4.2 below.

\subsection{Integration of the system (\ref{HH})}

The first three equations in (\ref{HH}) imply that $h_{33}=s h_3^{5/2}$, $s=const$. Thus, there are 2 cases to consider, $s=0$ and $s\ne 0$.  

\bigskip
 
\centerline {\bf Case I:  $s=0$}

\medskip

In this case  $h$ is linear in $u^3$ and the equations (\ref{HH}) imply
$$
h=\frac{u^3}{(c+au^1+bu^2)^2}+p(u^1, u^2).
$$
This ansatz can be simplified by utilizing  the canonical transformations \cite{Kup5},
\begin{equation}
\tilde u^1=\lambda u^1, ~~ \tilde u^2=u^2, ~~~ \tilde u^3=\frac{1}{\lambda}u^3,  ~~~ \tilde u^4=\frac{1}{\lambda^2}u^4, ...,
\label{can1}
\end{equation}
etc, $\lambda$=const, and
\begin{equation}
\tilde u^1=u^1, ~~ \tilde u^2=u^2+su^1, ~~~ \tilde u^3=u^3+2su^2+s^2u^1, ...,
\label {can2}
\end{equation}
etc, $s$=const.  Both transformations preserve the Poisson bracket
specified by (\ref{Ham}). Hence,  they can be used to  simplify  the Hamiltonian. Suppose, for instance, that $b\ne 0$. Then, up to the second canonical transformation, one can assume the ansatz
$h=u^3/(c+u^2)^2+p(u^1, u^2).$ If $b=0$ then $h=u^3/(c+u^1)^2+p(u^1, u^2).$ Thus, there are two subcases:

\medskip

\noindent {\bf Subcase ${\rm I}_1$: $h=u^3/(c+u^1)^2+p(u^1, u^2).$} The substitution of this ansatz into the four remaining equations for $h$ implies the following system for $p$:

\begin{eqnarray*}
&&p_{111}= -\frac{16(u^2)^2+(c+u^1)^3(8(u^2)^2p_{22}+4(3c+u^1)u^1p_{12}) +2(c+u^1)^4(5c+u^1)p_{11}}{c(c+u^1)^5(2+u^1(c+u^1)^2p_{22})} + \\
&&\frac{(c+u^1)(c^2p_{12}^2-c(c-3u^1)p_{22}p_{11})}{c(2+u^1(c+u^1)^2p_{22})}, \\ \\
&&p_{112}=2\frac{2u^2-c(c+u^1)^3p_{12}}{c(c+u^1)^4}, \quad
p_{122}=-\frac{2+c(c+u^1)^2p_{22}}{c(c+u^1)^3}, \quad
p_{222}=0.
\end{eqnarray*}

\noindent The last three equations lead to $p(u^1,u^2)= \frac{1+\alpha c(c+u^1)}{c(c+u^1)^2}(u^2)^2+\frac{\beta u^2}{c+u^1}+q(u^1)$ where $\alpha$ and $\beta$ are arbitrary constants. The substitution into the first equation gives a linear ODE for $q$,
\begin{eqnarray*}
(\alpha c^2 -3u^1\alpha c -4)q'' -(c+u^1)(1+c u^1 \alpha) q''' = \frac{c^2 \beta^2}{2(c+u^1)^3}.
\end{eqnarray*} 
Without any loss of generality one has
$$
q(u^1)=\frac{m}{(c+u^1)^2}+\frac{n}{c+u^1}
$$
where the constants $m, n$ satisfy a single relation $2(1-c^2\alpha)n-6c\alpha m+\frac{1}{2}c^2\beta^2=0$. Ultimately, we have Hamiltonian densities
$$
h=u^3/(c+u^1)^2+ \frac{1+\alpha c(c+u^1)}{c(c+u^1)^2}(u^2)^2+\frac{\beta u^2}{c+u^1}+\frac{m}{(c+u^1)^2}+\frac{n}{c+u^1}.
$$
Notice that the constant $\beta$ can be eliminated by the second canonical transformation. 

\medskip

\noindent {\bf Subcase ${\rm I}_2$: $h=u^3/(c+u^2)^2+p(u^1, u^2).$} The substitution of this ansatz into the four remaining equations for $h$ implies the following system for $p$,
\begin{eqnarray*}
&&p_{222}= \frac{(u^1)^3(c+u^2)p_{111}-2c(u^1)^2p_{11}-2cu^1(c-u^2)p_{12}-(5c-3u^2)(c-u^2)^2p_{22}}{(c-u^2)^3(c+u^2)},\\
&&p_{122}=\frac{(u^1)^2(c+u^2)p_{111}-2cu^1p_{11}-2(2c^2-3cu^2+(u^2)^2)p_{12}}{(c-u^2)^2(c+u^2)},\\
&&p_{112}=\frac{u^1(c+u^2)p_{111}+(u^2-3c)p_{11}}{(c-u^2)(c+u^2)},
\end{eqnarray*}
along with one more relation for $p_{111}$ which we do not write out due to its complexity. The first three relations imply $p(u^1, u^2)=\frac{u^1u^2-\alpha u^2-\frac{\alpha c}{3}+(c-u^2)^3g(\eta)}{(c+u^2)^2}$ where 
$g(\eta)$ is an arbitrary function of a single variable $\eta=\frac{u^1}{c-u^2}$. The result of the substitution of this ansatz into the remaining equation for $p_{111}$  factors into a product of two terms, leading to the cases (i) and (ii) below:

\noindent {\bf (i) } The function $g$ satisfies a first order ODE
$$
4c^2\eta ^2 g'-12c^2\eta g -1-\alpha \eta+2c\eta^2=0,
$$ 
the solution of this equation is $g(\eta) = \mu \eta^3 +\frac{\eta}{4c} -\frac{\alpha}{12c^2}-\frac{1}{16c^2\eta}.$ This results in the Hamiltonian densities of the form
$$
h=\frac{u^3}{(c+u^2)^2}+\frac{u^1u^2}{(c+u^2)^2}-\alpha \frac{u^2+c/3}{(c+u^2)^2}+\mu 
\frac{(u^1)^3}{(c+u^2)^2}+ \\ \frac{1}{4c}\frac{(c-u^2)^3}{(c+u^2)^2}\left(\frac{u^1}{c-u^2}-\frac{\alpha}{3c}
-\frac{c-u^2}{4cu^1}\right).
$$

\noindent {\bf (ii)} The function $g$ solves a third order ODE
$$
(4c^2\eta ^2 g'-12c^2\eta g -1-\alpha \eta+2c\eta^2)g''' +(\alpha-4c \eta +12c^2g +4c^2\eta g'-2c^2\eta ^2 g'')g''+ (4c-8c^2 g')g' -\frac{1}{2} =0.
$$
Remarkably, this complicated equation trivializes after being differentiated by $\eta$ once, taking the form
$$
(4c^2\eta ^2 g'-12c^2\eta g -1-\alpha \eta+2c\eta^2)g''''=0.
$$
Since the possibility when the coefficient at $g''''$ equals zero was considered above in the case (i), we
conclude that $g$ must be a cubic polynomial,
$$
g=\mu +\nu \eta + \gamma \eta^2+\delta \eta^3
$$
where the constants satisfy a single relation $12\delta-8c\nu+16c^2(\nu^2-3\gamma \mu)-4\gamma \alpha+1=0$ which can be obtained by substituting back into the original ODE. This leads to Hamiltonian densities of the form
$$
h=\frac{u^3}{(c+u^2)^2}+ \frac{u^1u^2-\alpha u^2-\frac{\alpha c}{3}+\mu(c-u^2)^3+\nu u^1(c-u^2)^2+\gamma (u^1)^2(c-u^2)+\delta(u^1)^3}{(c+u^2)^2}.
$$
Notice that the particular case  $c=0$ results (up to obvious equivalence transformations and relabeling of constants) in Hamiltonian densities of the form
\begin{equation}
h=\frac{u^3}{(u^2)^2}+\alpha \frac{(u^1)^2}{u^2}+\beta \frac{u^1}{u^2}+\gamma \frac{1}{u^2}+
\delta \frac{(u^1)^3}{(u^2)^2}
\label{lin2}
\end{equation}
where  $\alpha, \beta, \gamma, \delta$ satisfy a single relation $\beta^2-4\alpha \gamma +12 \delta=0$.

{\bf Remark.} The apparent similarity of the cases (\ref{lin}) and (\ref{lin2}) is not accidental and manifests  an important reciprocal invariance of the class of Hamiltonian chains (\ref{Ham}). Recall that the conservation  of momentum reads
\begin{eqnarray*}
&&u^2_t=(u^1h_1+2u^2h_2+3u^3h_3-h)_x.
\end{eqnarray*}
Let us change from $x, t$ to the new independent variables $X, T$ where 
$$
dX=u^2dx+(u^1h_1+2u^2h_2+3u^3h_3-h) dt, ~~~ T=t.
$$
It is known that reciprocal transformations of this type preserve Poisson brackets of the form (\ref{Ham}), see e.g. \cite{Fernon} (it is crucial that $u^2$ is the momentum of the corresponding Poisson bracket).
One can verify directly that performing the above change of independent variables and introducing
\begin{equation}
H=\frac{h}{u^2}, ~~ U^1=\frac{u^1}{u^2}, ~~ U^2=\frac{1}{u^2}, ~~ U^3=\frac{u^3}{(u^2)^3}, ~~ ..., ~~ U^n=\frac{u^n}{(u^2)^n},
\label{recipr}
\end{equation}
one arrives at the system which takes the original form (\ref{Ham}) in the variables $X, T, U^n, H$. Thus, the above reciprocal transformation is canonical. One can verify directly that the change of variables  (\ref{recipr}) indeed identifies (\ref{lin}) and (\ref{lin2}).

\bigskip

\centerline {\bf Case II: $s\ne 0$}

\medskip

In this case the  elementary integration  gives
$$
h=(\gamma u^3+p(u^1, u^2))^{1/3}+q(u^1, u^2),
$$
and the substitution into the last three equations (\ref{HH}) implies that $q$ is linear. Up to the  equivalence $h\to \alpha h +au^1+bu^2+c$  we thus have
$$
h=(u^3+p(u^1, u^2))^{1/3}.
$$
The substitution of this ansatz into the remaining equations for $h$ implies a complicated system of third order PDEs for $p(u^1, u^2)$. A useful observation is that this system is invariant under a 3-parameter group of point symmetries which is generated by the two canonical transformations (\ref{can1}), (\ref{can2}) and the reciprocal transformation (\ref{recipr}). 
The infinitesimal generators of these symmetries are 
$$
X_1=u^1\partial_{u^1}-p\partial_p, ~~~ X_2=u^1\partial_{u^2}-2u^2\partial_p, ~~~
X_3=(u^1)^2\partial_{u^1}+u^1u^2\partial_{u^2}+(3pu^1+2(u^2)^2)\partial_p;
$$
they satisfy the commutator relations
$$
[X_1, X_2]=X_2, ~~~ [X_1, X_3]=X_3, ~~~ [X_2, X_3]=0.
$$
These symmetries suggest a change of variables which considerably simplifies  the equations for $p$. The idea is to choose new coordinates such that the symmetry generators assume the simplest possible form. Introducing  $\xi=-\frac{1}{u^1}, \ \eta=\frac{u^2}{u^1}, \ s=\frac{pu^1+(u^2)^2}{(u^1)^4}$,
we have
$$
X_1=-\xi \partial_{\xi}-\eta \partial_{\eta}-4s \partial_{s}, ~~~ X_2= \partial_{\eta}, ~~~  X_3= \partial_{\xi}.
$$ 
Setting $s=s(\xi, \eta)$ we have $p=s(\xi, \eta)(u^1)^3-\frac{(u^2)^2}{u^1}$. In terms of $s(\xi, \eta)$ the equations for $p$ assume a remarkable symmetric form 
\begin{eqnarray}
&&s_{\xi \xi \eta}s_{\eta \eta}-s_{\eta \eta \eta}s_{\xi \xi}=4s_{\eta}, \label{s1} \\
&&s_{\xi \eta \eta}s_{\xi \xi}-s_{\xi \xi \xi}s_{\eta \eta}=4s_{\xi}, \label{s2} \\
&&s_{\eta \eta \eta}s_{\xi }+s_{\xi \eta \eta}s_{\eta }=s_{\eta \eta}s_{\xi \eta}, \label{s3} \\
&&s_{\xi \xi \xi}s_{\eta }+s_{\xi \xi \eta}s_{\xi }=s_{\xi \xi}s_{\xi \eta}, \label{s4} \\
&&s_{\eta \eta \eta}s_{\xi \xi}s_{\eta}+s_{\xi \eta \eta}s_{\eta \eta}s_{\xi}+4s_{\eta}^2=\frac{1}{2}s_{\eta \eta}(12s+s_{\xi \eta}^2+s_{\xi \xi} s_{\eta \eta}), \label{s5} \\
&&s_{\xi \xi \xi}s_{\eta \eta}s_{\xi}+s_{\xi \xi \eta}s_{\xi \xi}s_{\eta}+4s_{\xi}^2=\frac{1}{2}s_{\xi \xi}(12s+s_{\xi \eta}^2+s_{\xi \xi} s_{\eta \eta}); \label{s6} 
\end{eqnarray}
in the process of derivation of these equations we have assumed that $s_{\xi}, \ s_{\eta}$ and 
$s_{\xi \xi} s_{\eta}^2-s_{\eta \eta} s_{\xi}^2$ are nonzero: these expressions appear as denominators in the intermediate calculations. Particular cases when some of these expressions vanish will be discussed below. 

To solve the equations (\ref{s1}) -- (\ref{s6}) we proceed as follows. Differentiating (\ref{s3}),  (\ref{s4}) and using 
 (\ref{s1}), (\ref{s2}) we obtain four relations among the fourth order partial derivatives of $s$,
\begin{eqnarray*}
&&s_{\xi}s_{\eta \eta \eta \eta}+s_{\eta}s_{\xi \eta \eta \eta}=0, \quad 
s_{\xi}s_{\xi \eta \eta \eta}+s_{\eta}(s_{\xi \xi \eta \eta}-4)=0,  \\
&&s_{\xi}(s_{\xi \xi \eta \eta}-4)+s_{\eta}s_{\xi \xi \xi \eta}=0, \quad 
s_{\xi}s_{\xi \xi \xi \eta}+s_{\eta}s_{\xi \xi \xi \xi}=0,
\end{eqnarray*}
which can be parametrized as
\begin{equation}
 s_{\eta \eta \eta \eta}=q, ~~  s_{\xi \eta \eta \eta}=qr, ~~ s_{\xi \xi \eta \eta}-4=qr^2, ~~ s_{\xi \xi \xi \eta}=qr^3, ~~ s_{\xi \xi \xi \xi}=qr^4,
 \label{4der}
 \end{equation}
$r=-s_{\xi}/s_{\eta}$.     The further analysis leads to the two possibilities.

\noindent {\bf Subcase ${\rm II}_1$: $q=0$.} In this case $s(\xi, \eta)$ is a polynomial of the form
$$
s=\xi^2\eta^2+a\xi^3+b\xi^2\eta+c\xi\eta^2+d\eta^3+\alpha \xi^2+\beta \xi \eta +\gamma \eta^2+\mu \xi +\nu \eta +\epsilon.
$$
The substitution  into the remaining equations for $s$ implies the following relations among the coefficients:
$$
\beta=bc-9ad, ~~~ \mu=\alpha c-3a\gamma, ~~~ \nu=b\gamma-3\alpha d, ~~~  12\epsilon+\beta^2+4\alpha \gamma=4(b\nu+c\mu).
$$
Notice that the corresponding $p(u^1, u^2)=s(-\frac{1}{u^1}, \ \frac{u^2}{u^1})(u^1)^3-\frac{(u^2)^2}{u^1}$ will be a cubic polynomial in $u^1, u^2$. A particular example from this class with the  Hamiltonian density $h=(u^3+\tau)^{1/3}$  was discussed in  \cite{Kup3}. It corresponds to the case where $a=-\tau$ and all other coefficients of $s(\xi, \eta)$ are zero.

\noindent {\bf Subcase ${\rm II}_2$: $q\ne 0$.} Then the consistency conditions of (\ref{4der}) imply the relations
$q_{\xi}=(qr)_{\eta}, \ r_{\xi}=rr_{\eta}$. Taking into account that $r=-s_{\xi}/s_{\eta}$ we have
$s_{\xi \xi} s_{\eta}^2-s_{\eta \eta} s_{\xi}^2=0$. This case is discussed below. 

There are two more possibilities one needs to consider to complete the classification (notice that  the equations (\ref{s1}) -- (\ref{s6}) can no longer be used since they were derived under the assumption that certain expressions do not vanish). 

\noindent {\bf Subcase ${\rm II}_3$:} $s_{\xi}=0$ or $s_{\eta}=0$. A simple analysis leads to Hamiltonian densities of the form
$$
h =\left( u^3+\alpha (u^1)^3-\frac{(u^2)^2}{u^1}\right)^{1/3},
$$
$\alpha$ =const. 

\noindent {\bf Subcase ${\rm II}_4$:} $s_{\xi \xi} s_{\eta}^2-s_{\eta \eta} s_{\xi}^2=0$. Setting $r=-\frac{s_{\xi}}{s_{\eta}}$ one obtains
$$
r_{\xi}=r r_{\eta}, ~~~ s_{\xi}=-r s_{\eta}.
$$
Calculating partial derivatives of $s$ with the help of the above relations and substituting them into the conditions $H^1_{jk}=0$ we obtain $r_{\eta \eta}=0$; without any loss of generality one can set
$r=-\frac{\eta}{\xi}.$ This implies $\xi s_{\xi}-\eta s_{\eta}=0$, therefore, $s=s(z), \ z=\xi \eta=-u^2/(u^1)^2$. 
For $s(z)$ we obtain the ODE
$$
8z^2s'''s'+8zs''s'-4z^2(s'')^2-(s')^2-12s=0
$$
which linearizes after being differentiated by $z$ once. Its general solution is given by the formula
$$
s(z)=z^2+\alpha+\beta (-z)+\gamma (-z)^{1/2}+\delta (-z)^{3/2}
$$
where the constants $\alpha, \beta, \gamma, \delta$ satisfy a single quadratic relation
$12\alpha+\beta ^2 -3\gamma\delta=0.$
This results in  Hamiltonian densities of the form 
$$
h=\left(u^3+\alpha (u^1)^3+\beta u^1u^2+\gamma (u^1)^2(u^2)^{1/2}+\delta (u^2)^{3/2}\right)^{1/3}.
$$

\subsection{Generating functions for conservation laws}

In this section we demonstrate that any Hamiltonian  chain (\ref{Ham}) whose density $h$ satisfies the system (\ref{HH}), possesses a generating function which provides an infinity of conservation laws.
To illustrate the method of generating functions let us consider the Benney chain (\ref{Benney})
and introduce the generating function
\begin{equation}
\lambda=p+\frac{u^1}{p}+\frac{u^2}{p^2}+\frac{u^3}{p^3}+...
\label{gen}
\end{equation}
which was shown in \cite{Gibb81} to satisfy, by virtue of (\ref{Benney}),  the fundamental relation
\begin{equation}
\lambda_t-p\lambda_x=\lambda_p\left[p_t-(p^2/2+u^1)_x\right].
\label{fund}
\end{equation}
This relation provides an infinity of conserved densities $H_n({\bf u})$  defined by the equation
$$
p_t=(p^2/2+u^1)_x
$$
where one has to substitute the expression for $p(\lambda)$ obtained from (\ref{gen}): $p=\lambda-\frac{H_1}{\lambda}-\frac{H_2}{\lambda^2}-\frac{H_3}{\lambda^3}-...$. Explicitly, one gets
$H_1=u^1, \ H_2=u^2, \ H_3=u^3+(u^1)^2$, etc.

Given an arbitrary  Hamiltonian density $h(u^1, u^2, u^3)$, a  generating function is sought in the form
\begin{equation}
\lambda =\psi (p, u^{1}, u^{2})+\underset{k=1}{\overset{\infty }{\sum }}\frac{
u^{k}}{p^{k}};
\label{psi}
\end{equation}
notice that the dependence on higher moments $u^3, u^4, u^5, $ etc, is exactly the same as in (\ref{gen}). As demonstrated in \cite{Pavlov6}, the generating function (\ref{psi}) has to satisfy  the Gibbons-type relation
\begin{equation}
\lambda _{t}-(2ph_{3}+h_{2})\lambda _{x}=\lambda_p
\left[ p_{t}-(p^{2}h_{3}+ph_{2}+h_{1})_x\right],
\label{fund1}
\end{equation}
which reduces to (\ref{gen}) for the Benney Hamiltonian $h=(u^3+(u^1)^2)/2$. Substituting (\ref{psi}) into (\ref{fund1})  one obtains, by virtue of (\ref{Ham}),  the following relations among the first order partial derivatives  $\psi_1, \psi_2, \psi_p$:
\begin{eqnarray}
&&(\psi_p-\frac{3 u^3}{p^4})(p^2h_{13}+ph_{12}+h_{11})+\psi_1(2u^2h_{13}+u^1h_{12}-2ph_3)+\psi_2 (3u^3h_{13}+\nonumber\\
&&2u^2h_{12}+u^1h_{11})+\frac{1}{p^3}(3u^3h_{12}+(2u^2+\frac{3u^3}{p})h_{11})=0,\nonumber\\ 
\nonumber\\ 
&&(\psi_p-\frac{3 u^3}{p^4})(p^2h_{23}+ph_{22}+h_{12})+\psi_1(2u^2h_{23}+u^1h_{22}+2h_3)+\psi_2 (3u^3h_{23}+ \label{der}\\
&&2u^2h_{22}+u^1h_{12}-2ph_3)+\frac{1}{p^3}(3u^3h_{22}+(2u^2+\frac{3u^3}{p})h_{12})=0,\nonumber\\
&&(\psi_p-\frac{3 u^3}{p^4})(p^2h_{33}+ph_{23}+h_{13})+\psi_1(2u^2h_{33}+u^1h_{23})+\psi_2 (3u^3h_{33}+2u^2h_{23}+\nonumber\\
&&u^1h_{13}+2h_3)+\frac{1}{p^3}(3u^3h_{23}+(2u^2+\frac{3u^3}{p})h_{13}-2ph_3)=0.\nonumber
\end{eqnarray}
These relations uniquely specify  the partial derivatives $\psi_1, \psi_2, \psi_p$:
\begin{eqnarray}
& &\psi_1=\nonumber\\
& & -\frac{1}{\triangle}\left[ 4p^2h_3(ph_3h_{33}+h_{23}(2h_3+u^1h_{13})-h_{33}(u^1h_{12}+u^2h_{22}))\right.\nonumber\\ 
& &p(2h_3(3u^3(h_{33}h_{22}-h_{23}^2)+2h_3(h_{22}+h_{13})+u^1(h_{13}(2h_{22}+h_{13})-2h_{12}h_{23}-h_{11}h_{33}))+\nonumber\\
& &(u^1)^2(h_{12}(h_{12}+h_{33}-h_{13}h_{23})+h_{23}(h_{11}h_{23}-h_{12}h_{13})+h_{22}(h_{13}^2-h_{11}h_{33})))\nonumber\\
& &h_3(6u^3(h_{13}h_{23}-h_{12}h_{33})-2h_3h_{12}+u^1(h_{11}h_{23}-h_{12}h_{13}))+2u^2u^1(h_{12}(h_{12}h_{33}-h_{13}h_{23})+\nonumber\\
& &\left.h_{23}(h_{11}h_{23}-h_{12}h_{13})+h_{22}(h_{12}^2-h_{11}h_{33}))\right],\nonumber\\
\nonumber\\
& &\psi_2=\nonumber\\
& & -\frac{1}{\triangle}\left[ 4ph_3^2h_{33}p^2+4h_3(h_{23}(h_3+u^2h_{23}+u^1h_{13})-h_{33}(u^2h_{22}+u^1h_{12}))p+\right.\nonumber\\
& &(u^1)^2(h_{12}(h_{12}h_{33}-h_{13}h_{23})+h_{23}(h_{11}h_{23}-h_{12}h_{13})+h_{22}(h_{13}^2-h_{11}h_{33}))+\nonumber\\
& & \left.2h_{3}(h_{13}(2h_3+2u^2h_{23}+u^1(h_{13}+h_{22}))-h_{12}(2u^2h_{33}+u^1h_{23})-u^1h_{11}h_{33})\right],\nonumber\\
\nonumber\\
& &\psi_p=\nonumber\\
& & -\frac{1}{\triangle}\left[ 4h_3^2h_{33}u^1p^2+4h_3p(2h_3(u^2h_{33}+u^1h_{23})+u^2u^1(h_{23}^2-h_{22}h_{33})-(u^1)^2(h_{13}h_{23}-h_{12}h_{33}))+\right.\nonumber\\
& &2h_3(2h_3(3u^3h_{33}+4u^2h_{23}+u^1(2h_{13}+h_{22})+(u^1)^3(h_{12}(h_{12}h_{33}-h_{13}h_{23})+h_{23}(h_{23}h_{11}-h_{12}h_{13})+\nonumber\\
& &h_{22}(h_{13}^2-h_{11}h_{33})))+4(u^2)^2(h_{23}^2-h_{22}h_{33})+4u^1u^2(h_{13}h_{23}-h_{12}h_{33})-3u^1u^3(h_{23}^2-h_{22}h_{33})+\nonumber\\
& &\left.(u^1)^2(h_{13}(2h_{22}+h_{13})-2h_{23}h_{12}-h_{11}h_{33}))\right].\nonumber\\
\nonumber
\end{eqnarray}
\\
\begin{eqnarray}
&&\triangle= \nonumber\\
& &4p^4h_3^2h_{33}+4p^3h_2(2h_3h_{23}+u^2(h_{23}^2-h_{22}h_{33})+u^1(h_{13}h_{23}-h_{12}h_{33}))+((u^1)^2p^2+2u^2u^1 p)\nonumber\\
& &(2h_3(h_{13}(h_{13}+2h_{22})-2h_{12}h_{23}-h_{11}h_{33})+(h_{12}(h_{12}h_{33}-h_{13}h_{23})+h_{23}(h_{11}h_{23}-h_{12}h_{13})+\nonumber\\
& &h_{22}(h_{13}^2-h_{11}h_{33})))+2h_3p^2(3u^3(h_{22}h_{33}-h_{23}^2)+2h_3(2h_{13}+h_{22}))+4h_3p (3u^3(h_{13}h_{23}-h_{12}h_{33})-\nonumber\\
& &2h_3h_{12})+2h_{11}(h_3(2h_3+u^1h_{22}+4u^2h_{23})+2(u^2)^2(h_{23}^2-h_{22}h_{33}))-\nonumber\\
& &3u^3((h_{13}^2-h_{11}h_{33})(2h_3+u^1h_{22})+u^1h_{23}(h_{12}h_{33}-h_{13}h_{23})+u^1h_{23}(h_{11}h_{23}-h_{12}h_{13})).\nonumber
\end{eqnarray}
These equations are consistent provided the density $h$ satisfies the system (\ref{HH}). This proves Theorem 4: indeed, an infinite sequence of conservation laws results from the equation
$$
 p_{t}=(p^{2}h_{3}+ph_{2}+h_{1})_x
$$
 where one has to substitute the expansion for $p$ in terms of $\lambda$ obtained from (\ref{psi}). 

The  calculation of  $\psi (p, u^{1}, u^{2})$  can be summarized as follows:

\noindent --- one first integrates  the equation for $\psi_p$, which appears to be rational in $p$, with respect to $p$. This defines $\psi$ up to  a function of $u^1, u^2$.

\noindent --- one fixes this function  by a substitution  into the equations for $\psi_1, \psi_2$.

This procedure leads to the following explicit formulae for generating functions for the Hamiltonian densities obtained in Sect. 4.1.

\noindent {\bf Linear case.} This is the simplest generalization of the Benney Hamiltonian with the density
\begin{eqnarray*}
h(u^1,u^2,u^3)=u^3+\alpha (u^1)^2+\beta u^1u^2 +\gamma (u^2)^2+\delta (u^1)^3,\end{eqnarray*}
where the constants satisfy the condition $\beta ^2 -4\alpha \gamma+12\delta=0.$ Even in this case the  function $\psi$ is quite nontrivial,
\begin{eqnarray*}
\psi(p,u^1,u^2)=\frac{2}{(4\alpha\gamma-\beta^2)^{\frac{1}{2}}}\arctan 
\left[\frac{\beta+2\gamma p}{(1+\gamma u^1)(4\alpha\gamma-\beta^2)^ 
{\frac{1}{2}}}\right].
\end{eqnarray*}
The generating function is obtained by substituting this expression into (\ref{psi}).
\\
\\
\noindent {\bf Subcase ${\rm I}_1$.} In this case the Hamiltonian density takes the form
\begin{eqnarray*}
h(u^1,u^2,u^3)=\frac{u^3}{(c+u^1)^2}+\frac{1+\alpha c (c+u^1)}{c(c+u^1)^2}(u^2)^2+\frac{\beta u^2}{c+u^1}+
\frac{m}{(c+u^1)^2}+\frac{n}{c+u^1},
\end{eqnarray*}
where the constants satisfy the single relation, $2(1-c^2\alpha)n-6c\alpha m+\frac{1}{2}c^2\beta^2=0.$ The corresponding  function $\psi$ is given by
\begin{eqnarray*}
\psi(p,u^1,u^2)=-\frac{c^{\frac{3}{2}}}{(3m)^{\frac{1}{2}}}\tanh ^{-1}\left[\frac{2(c^2\alpha-1)u^2+c(2(c^2\alpha-1)p-\beta c)-c\beta u^1}{2(3cm)^{\frac{1}{2}}(1+c\alpha u^1)} \right].
\end{eqnarray*}
\\
\\
\noindent {\bf Subcase ${\rm I}_2$(i).} For the Hamiltonian density
\begin{eqnarray*}
h(u^1,u^2,u^3)=\frac{u^3+u^1u^2-\alpha (u^2+\frac{c}{3})+\mu (u^1)^3}{(c+u^2)^2}+\frac{(c-u^2)^3}{4c(c+u^2)^2}\left( \frac{u^1}{c-u^2}-\frac{\alpha}{3c}-\frac{c-u^2}{4c u^1}\right),
\end{eqnarray*}
the function $\psi $ is given by
\begin{eqnarray*}
\psi(p,u^1,u^2)=\frac{1}{(3\mu)^{\frac{1}{2}}}\arctan \left[\frac{(u^2)^2-2cu^2+4cpu^1+c^2}{4c(3\mu)^{\frac{1}{2}}(u^1)^2}\right].
\end{eqnarray*}
\\
\\
\noindent {\bf Subcase ${\rm II}_3$:} For the Hamiltonian density
\begin{eqnarray*}
h(u^1,u^2,u^3)=(u^3+\alpha (u^1)^3 -\frac{(u^2)^2}{u^1})^{\frac{1}{3}},
\end{eqnarray*}
the  function $\psi$ is given by
\begin{eqnarray*}
\psi(p,u^1,u^2)=\frac{1}{(3\alpha)^{\frac{1}{2}}}\arctan\left[\frac{p u^1-u^2}{(3\alpha)^{\frac{1}{2}}(u^1)^2}   \right].
\end{eqnarray*}
\\
Other examples obtained in Sect. 4.1 lead to more complicated expressions for $\psi$.

\section{Hydrodynamic reductions and the diagonalizability}

To illustrate the method of hydrodynamic reductions we consider the Benney chain (\ref{Benney}), 
\begin{eqnarray*}
&&u^1_t= u^2_x, \notag \\
&&u^2_t =u^3_x+u^1u^1_x, \notag \\
&&u^3_t= u^4_x+2u^2u^1_x, \notag \\
&&u^4_t =u^5_x+3u^3u^1_x, \notag 
\end{eqnarray*}
etc. Following the approach of \cite{GibTsa96, GibTsa99} let us seek solutions in the form
$u^i=u^i(R^1, \dots,  R^m)$ where the Riemann invariants $R^1, \dots,  R^m$ solve a diagonal system
$$
R^i_t=\lambda^i(R) R^i_x.
$$
Substituting this ansatz  into the Benney equations  and equating to zero coefficients at $R^i_x$  we arrive at the following relations:
\begin{eqnarray}
&&\partial_iu^2=\lambda^i\partial_i u, \label{u1} \\
&&\partial_iu^3=((\lambda^i)^2-u)\partial_i u, \label{u2} \\
&&\partial_iu^4=((\lambda^i)^3-u\lambda^i-2u^2)\partial_i u, \label{u3}  \\
&&\partial_iu^5=((\lambda^i)^4-u(\lambda^i)^2-2u^2\lambda^i-3u^3)\partial_i u, \label{u4} 
\end{eqnarray}
etc. Here $u= u^1, \ \partial_i=\partial_{R^i},\  i=1, ..., m$ (no summation!) The consistency conditions of the first three relations (\ref{u1})--(\ref{u3}) imply
$$
\partial_i\partial_ju=\frac{\partial_j\lambda^i}{\lambda^j-\lambda^i}\partial_iu+
\frac{\partial_i\lambda^j}{\lambda^i-\lambda^j}\partial_ju,
$$
$$
\partial_j\lambda^i\partial_iu+\partial_i\lambda^j\partial_ju=0,
$$
$$
\lambda^i\partial_j\lambda^i\partial_iu+\lambda^j\partial_i\lambda^j\partial_ju+\partial_iu\partial_ju=0,
$$
respectively. Solving these equations for $\partial_j\lambda^i$  we arrive at the Gibbons-Tsarev system
\begin{equation}
\partial_j\lambda^i=\frac{\partial_ju}{\lambda^j-\lambda^i}, ~~~
\partial_i\partial_ju=2\frac{\partial_iu\partial_ju}{(\lambda^i-\lambda^j)^2}.
\label{gib}
\end{equation}
It is a truly remarkable fact that all other consistency conditions (e.g., of the relation (\ref{u4}), etc), are satisfied identically modulo (\ref{gib}). Moreover, the semi-Hamiltonian property (\ref{semi}) is also automatically satisfied. Thus, the system (\ref{gib}) governs $m$-component reductions of the Benney chain. Up to reparametrizations $R^i\to f^i(R^i)$ these reductions depend on $m$ arbitrary functions of a single variable. Solutions arising within this approach are known as multiple waves, or  nonlinear interactions of planar simple waves.

The above approach clearly applies to any hydrodynamic chain from the class $C$. Let us restrict, for instance, to the chains of the type (\ref{max}). Looking for solutions in the form $u^i=u^i(R^1, \dots, R^m)$
and substituting this ansatz into (\ref{max}) we arrive at an infinite system of relations similar to (\ref{u1})--(\ref{u4}). The first three of them imply the `generalized Gibbons-Tsarev system' of the form
$$
\partial_j\lambda^i=(...)\partial_ju,  ~~~
\partial_i\partial_ju=(...)\partial_iu\partial_ju,
$$
$u=u^1$, where dots denote complicated expression which are {\it rational} in $\lambda^i$ with the coefficients depending on the chain under study (that is, on $g$, $h$, etc). Requiring that all  other consistency conditions, as well as the semi-Hamiltonian property, are satisfied  {\it identically}  we obtain  constraints for the matrix $V(u)$, see \cite{Fer3, Fer4, Fer5, Fer6, Fer7, Fer8} where a similar approach was applied to the classification of integrable multi-dimensional quasilinear systems. 
The main result of this section is the proof of the following theorem formulated in the introduction:

\medskip

\noindent {\bf Theorem 2} {\it
The vanishing of the Haantjes tensor is a necessary condition for the existence of  infinitely many hydrodynamic reductions and, thus, for the integrability of a  hydrodynamic chain.}

\medskip

\noindent We will give two different proofs of this statement. Based on essentially different ideas, they may be of interest in their own right. 

\medskip

\centerline{\bf  First Proof:}

\noindent This proof is computational. Writing down the equations of the chain in the form
$u^m_t=V^m_nu^n_x$ and substituting the ansatz $u^i=u^i(R^1, ..., R^m)$ we arrive at an infinite set of relations
$$
V^m_n\partial_iu^n=\lambda^i\partial_iu^m;
$$
we point out that all summations here and below involve finitely many nonzero terms.
Applying  the operator $\partial_j, \ j\ne i$, we obtain
\begin{equation}
V^m_{n,k}\partial_iu^n\partial_ju^k+V^m_n\partial_i\partial_j u^n=\partial_j\lambda^i \partial_iu^m+\lambda^i \partial_i\partial_j u^m.
\label{V}
\end{equation}
Interchanging the indices $i$ and $j$ and subtracting the results we arrive at the expression for 
$\partial_i\partial_j u^m$ in the form
$$
\partial_i\partial_j u^m=\frac{\partial_j\lambda^i}{\lambda^j-\lambda^i}\partial_iu^m
+\frac{\partial_i\lambda^j}{\lambda^i-\lambda^j}\partial_ju^m
+\frac{V^m_{n,k}-V^m_{k,n}}{\lambda^i-\lambda^j}\partial_iu^n\partial_ju^k.
$$
Substituting this back into (\ref{V}) we arrive at a simple relation
$$
 \partial_j\lambda^i \partial_iu^m+ \partial_i\lambda^j \partial_ju^m=\frac{N^m_{nk} \partial_iu^n\partial_ju^k}{\lambda^i-\lambda^j}
 $$
 where $N$ is the Nijenhuis tensor of $V$. This  can be rewritten in the invariant form
 \begin{equation}
 \partial_j\lambda^i \partial_i{\bf u}+ \partial_i\lambda^j \partial_j{\bf u}=\frac{N(\partial_i{\bf u}, \ \partial_j{\bf u})}{\lambda^i-\lambda^j}
 \label{NN}
 \end{equation}
which implies the following four relations:
$$
\begin{array}{c}
(\lambda^i)^2 \partial_j\lambda^i \partial_i{\bf u}+ (\lambda^j)^2\partial_i\lambda^j \partial_j{\bf u}=\frac{V^2N(\partial_i{\bf u}, \ \partial_j{\bf u})}{\lambda^i-\lambda^j}, \\
\ \\
(\lambda^i)^2 \partial_j\lambda^i \partial_i{\bf u}+ \lambda^i\lambda^j\partial_i\lambda^j \partial_j{\bf u}=\frac{VN(V\partial_i{\bf u}, \ \partial_j{\bf u})}{\lambda^i-\lambda^j}, \\
\ \\
\lambda^i\lambda^j \partial_j\lambda^i \partial_i{\bf u}+ (\lambda^j)^2\partial_i\lambda^j \partial_j{\bf u}=\frac{VN(\partial_i{\bf u}, \ V\partial_j{\bf u})}{\lambda^i-\lambda^j}, \\
\ \\
\lambda^i\lambda^j \partial_j\lambda^i \partial_i{\bf u}+ \lambda^i \lambda^j\partial_i\lambda^j \partial_j{\bf u}=\frac{N(V\partial_i{\bf u}, \ V\partial_j{\bf u})}{\lambda^i-\lambda^j}.
\end{array}
$$
For instance, the first relation can be obtained by applying the operator $V^2$ to  (\ref{NN}) and 
using $V\partial_i{\bf u}=\lambda^i\partial_i{\bf u}$.  Thus,
$$
V^2N(\partial_i{\bf u}, \ \partial_j{\bf u})-VN(V\partial_i{\bf u}, \ \partial_j{\bf u})-VN(\partial_i{\bf u}, \ V\partial_j{\bf u})+N(V\partial_i{\bf u}, \ V\partial_j{\bf u})=0.
$$
The last relation can be rewritten in the form $H(\partial_i{\bf u}, \ \partial_j{\bf u})=0$ where $H$ is the Haantjes tensor, indeed, a coordinate-free form of the relation (\ref{H}) is
$$
H(X, Y)=V^2N(X, Y)-VN(VX, Y)-VN(X,  VY)+N(VX,  VY)
$$
where $X, Y$ are arbitrary vector fields. Keeping in mind that
 (a)  $\partial_i{\bf u}$ and $ \partial_j{\bf u}$ are eigenvectors of the matrix $V$ corresponding to the eigenvalues $\lambda^i$ and $\lambda^j$, and 
 (b) $\lambda^i$ and $\lambda^j$ can take arbitrary values, we conclude that $H(X, Y)=0$ for any two formal eigenvectors of the matrix $V$.

Assuming that formal eigenvectors of  $V$
span the space  of dependent  variables ${\bf u}$
 (this is true for all examples discussed in this paper), we obtain $H=0$. In more detail,  let $X(\lambda)=(\xi^1(\lambda),\  \xi^2(\lambda),\  \xi^3(\lambda), ...)^t$ be a formal eigenvector of the matrix $V$ corresponding to the eigenvalue $\lambda$.   Let us assume that these eigenvectors span the space of dependent variables, that is, that there exist no non-trivial relations of the form $c_i\xi^i(\lambda)=0$ with finitely many nonzero $\lambda$-independent coefficients $c_i$. In other words, $\xi^i(\lambda)$ are linearly independent as polynomials in $\lambda$.

 The condition
 $H(X(\lambda), X(\mu))=0$, written in components, takes the form $H^i_{jk}\xi^j(\lambda)\xi^k(\mu)=0$;
 recall that, since the matrix $V$ belongs to the class $C$,  these sums contain finitely many terms for any fixed value of the upper index $i$. Taking into account the linear independence of $\xi^j(\lambda)$ and $\xi^k(\mu)$, one readily arrives at $H^i_{jk}=0$. As an illustration, let $V$ be the matrix corresponding to the Benney chain
 (\ref{Benney}). Then $X(\lambda)=(1, \ \lambda,\  \lambda^2-u^1,\ \lambda^3-u^1\lambda-2u^2, ...)^t$, so that each $\xi^i(\lambda)$ is a polynomial in $\lambda$ of the degree $i-1$. These eigenvectors  span the space of dependent variables since polynomials $\xi^i(\lambda)$ are manifestly linearly independent. 

Notice that we have proved a more general result, namely, that the existence of sufficiently many {\it two-component} reductions already implies the vanishing of the Haantjes tensor $H$. Indeed, nothing changes in the proof if we set $i=1, \ j=2$ in the formula (\ref{NN}). As demonstrated in the Theorem 3 below,  one can further strengthen the result by reversing the above proof under the additional assumption of the simplicity of the spectrum of the matrix $V$. This will require the  relations
$$
 \partial_i\lambda^j \partial_j{\bf u}=\frac{N(V \partial_i{\bf u}, \ \partial_j{\bf u})-VN(\partial_i{\bf u}, \ \partial_j{\bf u})}{(\lambda^i-\lambda^j)^2}, ~~~
  \partial_j\lambda^i \partial_i{\bf u}=\frac{N(V \partial_j{\bf u}, \ \partial_i{\bf u})-VN(\partial_j{\bf u}, \ \partial_i{\bf u})}{(\lambda^i-\lambda^j)^2}, 
$$
which can be obtained by applying $V$ to both sides of (\ref{NN}) and solving for 
$ \partial_i\lambda^j \partial_j{\bf u}$ and $  \partial_j\lambda^i \partial_i{\bf u}$.

\medskip

\centerline{\bf  Second Proof:}

\noindent Our first remark is that for  the chains from the class C one needs to know only {\it finitely many} rows of the matrix $V(u)$  to calculate each particular component of the Haantjes tensor. Let us fix the values of indices $i, j, k$ and denote  by $C(i, j, k)$ the maximal number of rows needed to calculate $H^i_{jk}$ (counting from the first row). We need to show that $H^i_{jk}=0$. Let us consider  an $m$-component diagonal reduction $u^i(R^1, \dots, R^m), \ i=1, 2, ...$. Choosing the first $m$ variables $u^1, \dots, u^m$ as independent, we can represent the reduction  explicitly as
$$
u^{m+1}=u^{m+1}(u^1, \dots, u^m), ~~~ u^{m+2}=u^{m+2}(u^1, \dots, u^m), \dots, 
$$
etc. Substituting these expressions into the first $m$ equations of the chain we obtain an $m$-component  system $S_m$ for $u^1, \dots, u^m$, while the remaining equations  will be satisfied identically (by  the definition of a reduction). Notice that the Haantjes tensor of the reduced system $S_m$
is identically zero since the reduction is diagonalizable. Let us now choose the number $m$ sufficiently large so that the first $C(i, j, k)$ equations of the chain  do not contain variables $u^{m+1}, u^{m+2}, \dots$ (one can always do so since any equation of the chain depends on finitely many $u$'s, and $m$ can be arbitrarily large). Then the first $C(i, j, k)$ equations of the reduced system $S_m$ will be
{\it identical} to the first $C(i, j, k)$ equations of the original infinite chain. Hence, the corresponding components $H^i_{jk}$ for the reduced system and for the infinite chain will also coincide. This proves that all components of the Haantjes tensor of the chain must be zero.

\medskip

A straightforward modification of the second proof allows one to show that the existence of an infinity of semi-Hamiltonian reductions implies the vanishing of the tensor $P$. This establishes the necessity  of the conjecture formulated in the Introduction.  

We emphasize that the condition of diagonalizability  alone is not sufficient for the integrability in general. This can be seen as follows.

{\bf Example.} Let us consider the chain
\begin{eqnarray*}
&&u^1_t= u^2_x +p(u^1)u^1_x, \notag \\
&&u^2_t =u^3_x+p(u^1)u^2_x+u^1u^1_x, \notag \\
&&u^3_t= u^4_x+p(u^1)u^3_x+2u^2u^1_x, \notag \\
&&u^4_t =u^5_x+p(u^1)u^4_x+3u^3u^1_x, \notag 
\end{eqnarray*}
etc, which is obtained from the Benney chain ${\bf u}_t=V({\bf u}) {\bf u}_x$ by the transformation
$V\to V+p(u^1)E$ where $E$ is an infinite identity matrix and $p$ is a function of $u^1$. One can  verify that the corresponding Haantjes tensor is  zero (which is not at all surprising since the addition of a multiple of the identity does not effect the diagonalizability). 
A simple calculation shows that hydrodynamic reductions of this chain are governed by exactly the same equations as in the Benney case, the only difference is that now the Riemann invariants $R^i$ solve the equations
$$
R^i_t=(\lambda^i(R)+p(u^1))R^i_x.
$$
The semi-Hamiltonian property is satisfied if and only if $p''=0$. Thus, we have constructed examples  which possess infinitely many diagonal hydrodynamic reductions none of which are semi-Hamiltonian
(if $p''\ne 0$).

\medskip

Theorem 2 can be strengthened under the additional assumption of  the simplicity of the spectrum of $V$. 

\medskip

\noindent {\bf Theorem 3} {\it
In the  simple spectrum case the vanishing of the Haantjes tensor $H$ is  necessary and sufficient  for the existence of two-component reductions parametrized by two arbitrary functions of a single argument.
}

\medskip

\centerline{\bf  Proof:}

\noindent The necessity part is contained in the first proof of Theorem 2. To establish the sufficiency one has to show that the vanishing of the Haantjes tensor implies the solvability of the equations
\begin{equation}
V\partial_1{\bf u}=\lambda^1\partial_1{\bf u}, ~~~ V\partial_2{\bf u}=\lambda^2\partial_2{\bf u},
\label{PP1}
\end{equation}
\begin{equation}
\partial_1\partial_2 u^m=\frac{\partial_2\lambda^1}{\lambda^2-\lambda^1}\partial_1u^m
+\frac{\partial_1\lambda^2}{\lambda^1-\lambda^2}\partial_2u^m
+\frac{V^m_{n,k}-V^m_{k,n}}{\lambda^1-\lambda^2}\partial_1u^n\partial_2u^k
\label{PP2}
\end{equation}
and
\begin{equation}
 \begin{array}{c}
 \partial_1\lambda^2 \partial_2{\bf u}=\frac{N(V \partial_1{\bf u}, \ \partial_2{\bf u})-VN(\partial_1{\bf u}, \ \partial_2{\bf u})}{(\lambda^1-\lambda^2)^2}, \\
 \ \\
  \partial_2\lambda^1 \partial_1{\bf u}=\frac{N(V \partial_2{\bf u}, \ \partial_1{\bf u})-VN(\partial_2{\bf u}, \ \partial_1{\bf u})}{(\lambda^1-\lambda^2)^2}, 
  \end{array}
\label{PP3}
\end{equation}
which govern two-component reductions. Our first observation is that the vanishing of the Haantjes tensor implies that the vectors $N(V \partial_1{\bf u}, \ \partial_2{\bf u})-VN(\partial_1{\bf u}, \ \partial_2{\bf u})$ and $N(V \partial_2{\bf u}, \ \partial_1{\bf u})-VN(\partial_2{\bf u}, \ \partial_1{\bf u})$ are the eigenvectors of $V$ with the eigenvalues $\lambda^2$ and $\lambda^1$, respectively. By the assumption of  the simplicity of the spectrum, they are proportional to $ \partial_2{\bf u}$ and $ \partial_1{\bf u}$. Thus,  equations (\ref{PP3}) reduce to a pair of first order PDEs for $\lambda^1$ and $\lambda^2$,
\begin{equation}
 \partial_1\lambda^2=\frac{k_1}{(\lambda^1-\lambda^2)^2}, ~~~  \partial_2\lambda^1=\frac{k_2}{(\lambda^1-\lambda^2)^2},
 \label{PP4}
 \end{equation}
 here $k_1$ and $k_2$ are the corresponding coefficients of proportionality. The  relations (\ref{PP1}) allow one to reconstruct all components $u^2, u^3, ...$ of the infinite vector ${\bf u}$ in terms of its first component $u^1$.
 Finally,   the equations (\ref{PP2}), which are nothing but the consistency conditions of (\ref{PP1}),  reduce to a single second order PDE for the first component $u^1$ of the infinite vector ${\bf u}$. 
Thus,  relations (\ref{PP1}) -- (\ref{PP3}) reduce to a pair of first order equations
(\ref{PP4}) plus one second order PDE for $u^1$. Up to reparametrizations $R^1\to f^1(R^1), \ 
R^2\to f^2(R^2)$, their general solution  depends on two arbitrary functions of a single variable.

\section{Conclusion}

We have proposed a simple and easy-to-verify necessary condition for the integrability of hydrodynamic chains based on the vanishing of the Haantjes tensor. Conservative and Hamiltonian examples are discussed,  illustrating the general approach. We conjecture that all examples arising in Sect. 3 and 4 from the requirement of the vanishing of the Haantjes tensor (and, thus, satisfying a necessary condition for the integrability) also satisfy the following  properties:

\noindent (i) they possess infinitely many $m$-component hydrodynamic reductions parametrized by $m$ arbitrary functions of a single variable (thus, they are integrable in the sense of  Definition 3); 

\noindent (ii) they belong to infinite hierarchies of commuting hydrodynamic chains. 

Our method leads to an  abundance of new  examples of hydrodynamic chains which require a further detailed investigation. We will address these issues in the future. 

We hope that the results of this paper will find important applications in the theory of infinite-dimensional Frobenius manifolds  (yet to be constructed).

\section*{Appendix 1:  the invariant  formulation of  the semi-Hamiltonian property}

For $m$-component systems (\ref{1d}) there exists a  tensor object which is responsible for the semi-Hamiltonian property. First of all one  computes  $(1, 3)$-tensors $M$ and $K$,
$$
M^s_{kij}=N^s_{kp}v^p_qN^q_{ij}+N^s_{pq}v^p_kN^q_{ij}-N^s_{pq}N^p_{ik}v^q_j
-N^s_{pq}N^p_{kj}v^q_i-N^s_{kp}N^p_{iq}v^q_j-N^s_{kp}N^p_{qj}v^q_i
$$
and
$$
\begin{array}{c}
K^s_{kij}=b^s_p\partial_{u^k}N^p_{ij}-b^p_k\partial_{u^p}N^s_{ij}+N^p_{ij}\partial_{u^p}b^s_k-
N^s_{kp}\partial_{u^i}b^p_j+N^s_{kp}\partial_{u^j}b^p_i \\
\ \\
+b^s_p\partial_{u^i}N^p_{jk}-b^p_i\partial_{u^p}N^s_{jk}+N^p_{jk}\partial_{u^p}b^s_i-
N^s_{ip}\partial_{u^j}b^p_k+N^s_{ip}\partial_{u^k}b^p_j \\
\ \\
+b^s_p\partial_{u^j}N^p_{ki}-b^p_j\partial_{u^p}N^s_{ki}+N^p_{ki}\partial_{u^p}b^s_j-
N^s_{jp}\partial_{u^k}b^p_i+N^s_{jp}\partial_{u^i}b^p_k;
\end{array}
$$
here $b=v^2$, that is, $b^i_j=v^i_pv^p_j$. Using $M$ and $K$ one defines a $(1, 3)$-tensor $Q$ as
$$
\begin{array}{c}
Q^s_{kij}=v^p_kK^s_{pqj}v^q_i+v^p_kK^s_{piq}v^q_j-v^p_qv^q_kK^s_{pij}-K^s_{kpq}v^p_iv^q_j\\
\ \\
+4v^p_kM^s_{pij}-2M^s_{kpj}v^p_i-2M^s_{kip}v^p_j.
\end{array}
$$
Ultimately, one introduces a tensor $P$,
\begin{equation}
P^s_{kij}=v^s_pQ^p_{kqj}v^q_i+v^s_pQ^p_{kiq}v^q_j-
v^s_qv^q_pQ^p_{kij}-Q^s_{kpq}v^p_iv^q_j.
\label{P}
\end{equation}
\begin{theorem} \cite{Pavlov3}
 A  diagonalizable  system (\ref{1d})  is semi-Hamiltonian if and only if the   tensor $P$ vanishes identically. 
\end{theorem}
These objects can be  calculated using computer algebra. Notice that they are well-defined for hydrodynamic chains from the class $C$: all tensor operations will involve finite summations only. 
Invariant coordinate-free definitions of the above tensors can be found in \cite{Pavlov3}.

\medskip

It was pointed out in \cite{Sevennec} (Proposition 5, p.19) that any strictly hyperbolic conservative system
\begin{equation}
u^i_t=f^i(u)_x, ~~~ i=1,..., m
\label{ap}
\end{equation}
with the zero Haantjes tensor is necessarily semi-Hamiltonian. The shortest  proof of this statement known to us can be summarized as  follows. Let us first rewrite (\ref{ap}) in the  Riemann invariants,  $R^i_t=\lambda^i(R)R^i_x$,
where $\lambda^i\ne \lambda^j$ due to the strict hyperbolicity. The conserved densities $u(R)$ satisfy an over-determined system of second order linear PDEs \cite{Tsarev},
\begin{equation}
\partial_i\partial_ju=\frac{\partial_j\lambda^i}{\lambda^j-\lambda^i}\partial_iu+
\frac{\partial_i\lambda^j}{\lambda^i-\lambda^j}\partial_ju, ~~~ i\ne j.
\label{22}
\end{equation}
The consistency conditions $\partial_k(\partial_i\partial_ju)=\partial_j(\partial_i\partial_ku), \  i\ne j\ne k,$
lead to linear relations among the first order derivatives of $u$ of the form $\sum c_n\partial_nu=0$ where  $c_n$ are certain functions of  $\lambda$'s. It is well-known that the vanishing of all coefficients $c_n$, that is, the involutivity of the linear system (\ref{22}), is equivalent to the semi-Hamiltonian property (\ref{semi}). In this case we have infinitely many conserved densities parametrized by $m$ arbitrary functions of  a single variable. It turns out that the requirement of  the existence of $m$ {\it functionally independent} solutions of the linear system (\ref{22}) is already  sufficient to conclude that all coefficients $c_n$ must be zero. Indeed, if a relation of the form $\sum c_n\partial_nu=0$ is satisfied by m
functionally independent $u$'s, the vector $c_n$ will be a zero eigenvector of the corresponding Jacobian matrix which is non-degenerate. Thus, $c_n=0$ and we have shown that
$$
(m \ {\rm conservation\  laws})+({\rm Riemann \ invariants})\Rightarrow (\infty {\rm \ of \ conservation \ laws}).
$$

\medskip

It would be of interest to obtain a direct tensor proof of the above result by showing that the vanishing of the Haantjes tensor $H$ for a conservative system (\ref{ap}) implies the vanishing of the tensor $P$. Such a proof would then generalize to hydrodynamic chains, for which we have yet no rigorous definition of Riemann invariants.

\section*{Appendix 2:  higher conservation laws for Hamiltonian chains}

Let us assume that there exists a conservation law of the form 
$$
P(u^1,u^2,u^3,u^4)_t = Q(u^1,u^2,u^3,u^4,u^5)_x.
$$
\noindent Using (\ref{Ham}) and collecting coefficients at $u^i_x$ we  obtain the expressions for $Q_i=\partial Q/\partial_{u^i}$, 
\begin{eqnarray*}
Q_1&=&5u^5P_4h_{13} + \sum_{i=1}^2(4u^4P_{5-i}h_{1,(i+1)}+u^1P_{3-i}h_{1,i}) +\sum_{j=1}^3(3u^3P_{5-j}h_{1,j}+2u^2P_{4-j}h_{1,j})+\\
& & P_1h_2,\\
Q_2&=&5u^3P_4h_{23} + \sum_{i=1}^2(4u^4P_{5-i}h_{2,(i+1)}+u^1P_{3-i}h_{2,i}) +\sum_{j=1}^3(3u^3P_{5-j}h_{2,j}+2u^2P_{4-j}h_{2,j})+\\
& &P_2h_2+2P_1h_3,\\
Q_3&=&5u^5P_4h_{33} + \sum_{i=1}^2(4u^4P_{5-i}h_{3,(i+1)}+u^1P_{3-i}h_{3,i}) +\sum_{j=1}^3(3u^3P_{5-j}h_{3,j}+2u^2P_{4-j}h_{3j})+\\
& &P_3h_2+2P_2h_3,\\
Q_4&=&h_2P_4+2h_3P_3, \\
Q_5&=&2h_3P_4.
\end{eqnarray*}

\noindent The consistency conditions  $Q_{ij}=Q_{ji}$ imply the following expressions for second order partial derivatives of $P$,
 
 \begin{eqnarray*}
 P_{11}&=&\frac{6u^4P_4h_{13}^2+9u^3P_4h_{12}h_{13}+3u^2P_4(h_{12}^2+2h_{13}h_{11})+3u^1P_4h_{11}h_{12}+2P_3h_3h_{11}}{2h_3^2},\\
 P_{12}&=& \frac{12u^4P_4h_{23}h_{13}+9u^3P_4(h_{22}h_{13}+h_{23}h_{12})+6u^2P_4(h_{12}(h_{22}+h_{13})+h_{23}h_{11})}{4h_3^2}+\\
& & \frac{3u^1P_4(h_{12}^2+h_{22}h_{11})+2h_3(3p_4h_{11}+2p_3h_{12})}{4h_3^2},\\
P_{13}&=&\frac{12u^4P_4h_{13}h_{33}+9u^3P_4(h_{23}h_{13}+h_{33}h_{12})+6u^2P_4(h_{13}^2+h_{23}h_{12}+h_{33}h_{11})}{4h_3^2}+\\
& & \frac{3u^1P_4(h_{13}h_{12}+h_{23}h_{11})+2h_3(3P_4h_{12}+2P_3h_{13})}{4h_3^2},\\
P_{14}&=&\frac{3P_4h_{13}}{2h_3},\\
P_{22}&=&\frac{6u^4P_4h_{23}^2+9u^3P_4h_{22}h_{23}+3u^2P_4(h_{22}^2+2h_{23}h_{12})+3u^1P_4h_{12}h_{22}+2h_3(3P_4h_{12}+P_3h_{22})}{2h_3^2},\\
 P_{23}&=& \frac{12u^4P_4h_{23}h_{33}+9u^3P_4(h_{23}^2+h_{22}h_{33})+6u^2P_4(h_{23}(h_{22}+h_{13})+h_{33}h_{12})}{4h_3^2}+\\
& & \frac{3u^1P_4(h_{13}h_{22}+h_{23}h_{12})+2h_3(3P_4h_{13}+2P_3h_{23}+3P_4h_{22})}{4h_3^2},\\
P_{24}&=&\frac{3P_4h_{23}}{2h_3},\\
P_{33}&=&\frac{6u^4P_4h_{33}^2+3P_4(3u^3h_{33}h_{23}+u^2(h_{23}^2+2h_{33}h_{13})+u^1h_{23}h_{13})+2h_{3}(3P_4h_{23}+P_3h_{33})}{2h_3^2},\\
P_{34}&=&\frac{3P_4h_{33}}{2h_3},\\
P_{44}&=&0.
\end{eqnarray*}
 
\noindent Upon ensuring the consistency of this system we obtain the set of relations (\ref{HH}).  We believe that the same process applied to any higher order conservation law
$$
P(u^1,\dots,u^m)_t=Q(u^1,\dots,u^{m+1})_x
$$
 will results in the same relations (\ref{HH}).

\section*{Acknowledgements}

We thank Karima Khusnutdinova for providing us with a  program for computing components of the Haantjes tensor and a further help with computer algebra.  We also thank John Gibbons, Boris Kupershmidt and Maxim Pavlov for  their interest and clarifying discussions, and the referee for useful comments.

The research of EVF was partially supported by the
European Union through the FP6 Marie Curie RTN project ENIGMA (Contract
number MRTN-CT-2004-5652), the ESF programme MISGAM and the EPSRC grant EP/D036178/1.


\begin{thebibliography}{99}

\bibitem{Alber} S.J. Alber, Associated integrable systems, J. Math. Phys. {\bf 32} (1991) 916-922.

\bibitem{Benney} D.J. Benney, Some properties of long nonlinear waves, Stud. Appl. Math. {\bf 52} (1973) 45-50.

\bibitem{Bla1} M. Blaszak,  Classical $R$-matrices on Poisson algebras and related dispersionless systems. Phys. Lett. A {\bf 297}, no. 3-4 (2002) 191-195. 

\bibitem{Buk} V.M. Bukhshtaber, D.V. Leikin and M.V. Pavlov, Egorov hydrodynamic chains, the Chazy equation, and the group ${\rm SL} (2, C)$, Funktsional. Anal. i Prilozhen. {\bf 37}, no. 4 (2003) 13-26.

%\bibitem{Burnat1} M. Burnat, The method of Riemann invariants for
%multi-dimensional nonelliptic system, Bull. Acad. Polon. Sci. Sr.
%Sci. Tech. {\bf 17} (1969) 1019-1026.


%\bibitem{Burnat3} M. Burnat, The method of characteristics and
%Riemann's invariants for multidimensional hyperbolic systems,
%(Russian) Sibirsk. Mat.  Z. {\bf 11} (1970) 279-309.

\bibitem{Chazy} J. Chazy, Sur les \'equations diff\'erentielles dont  l'int\'egrale g\'en\'erale poss
\`ede un coupure essentielle mobile, C.R. Acad. Sci. Paris {\bf 150} (1910) 456-458.
  
\bibitem{Dub} B.A. Dubrovin and S.P. Novikov, Hydrodynamics of weakly
deformed soliton lattices:
differential geometry and Hamiltonian theory, Russian Math. Surveys,
{\bf 44} , no. 6 (1989) 35-124.
 
 \bibitem{Dubr} B.A. Dubrovin, Geometry of 2D topological field
theories, Lect. Notes in Math.
{\bf 1620}, Springer-Verlag (1996) 120-348.

\bibitem{DubZ} B.A. Dubrovin, S. Liu and Y. Zhang, On Hamiltonian perturbations of hyperbolic systems of conservation laws. I. Quasi-triviality of bi-Hamiltonian perturbations, Comm. Pure Appl. Math. {\bf 59}, no. 4 (2006)  559-615.
 

\bibitem{Fernon} E.V. Ferapontov, Nonlocal Hamiltonian operators of hydrodynamic
type: differential geometry and applications, Amer. Math. Soc. Transl. (2),
{\bf 170} (1995) 33-58.

%\bibitem{Fer} E.V. Ferapontov,  D.A. Korotkin  and V.A. Shramchenko,
%Boyer-Finley equation and systems of hydrodynamic type, Class.
%Quantum Grav. {\bf 19} (2002) L1-L6.

%\bibitem{Fer2} E.V. Ferapontov and M.V. Pavlov, Hydrodynamic
%reductions of the heavenly equation, Class. Quantum Grav. {\bf 20}
%(2003) 1-13.


\bibitem{Fer3} E.V. Ferapontov and K.R. Khusnutdinova, On integrability of (2+1)-dimensional quasilinear systems, Comm. Math. Phys. {\bf 248} (2004) 187-206; arXiv:nlin.SI/0305044.

\bibitem{Fer4} E.V. Ferapontov and K.R. Khusnutdinova, The characterization of two-component (2+1)-dimensional integrable  systems of hydrodynamic type,   J. Phys. A: Math. Gen. {\bf 37}, N8 (2004)
2949 - 2963; arXiv:nlin.SI/0310021.

\bibitem{Fer5} E.V. Ferapontov and K.R. Khusnutdinova, Hydrodynamic reductions of multi-dimensional dispersionless PDEs: the test for integrability,   J. Math. Phys. {\bf 45}, N6 (2004) 2365-2377;  arXiv:nlin.SI/0312015.

\bibitem{Fer6} E.V. Ferapontov and K.R. Khusnutdinova, Double waves in multi-dimensional systems of hydrodynamic type: the necessary condition for integrability,  Proc. R. Soc.  A  {\bf 462} (2006) 1197-1219; arXiv:nlin.SI/0412064.

\bibitem{Fer7}  E.V. Ferapontov,  K.R. Khusnutdinova and M.V. Pavlov, Classification of integrable 
(2+1)-dimensional quasilinear hierarchies, Theor. Math. Phys.  {\bf 144} (2005)  35-43.

\bibitem{Fer77} E.V. Ferapontov, K.R. Khusnutdinova, D.G. Marshall and M.V. Pavlov,
Classification of integrable Hamiltonian hydrodynamic chains associated with Kupershmidt's brackets,
J. Math. Phys. {\bf 47}, no.1 (2006).

\bibitem{Fer8} E.V. Ferapontov,  K.R. Khusnutdinova and S.P. Tsarev, On a class of three-dimensional integrable Lagrangians,  Comm. Math. Phys. {\bf 261}, N1 (2006)  225-243.


\bibitem{Gibb81} J. Gibbons, Collisionless Boltzmann equations and
integrable moment equations, Physica {\bf 3D}  (1981) 503-511.

\bibitem{Gibb94} J. Gibbons and Y. Kodama,   A method for solving the
dispersionless KP hierarchy and its exact solutions. II, Phys. Lett.
A {\bf 135}, no. 3 (1989) 167--170.

\bibitem{GibTsa96} J. Gibbons and S. P. Tsarev, Reductions of the
Benney equations, Phys. Lett. A {\bf 211} (1996) 19-24.

\bibitem{GibTsa99} J. Gibbons and S. P. Tsarev, Conformal maps and
reductions of the Benney equations, Phys. Lett. A {\bf 258} (1999)
263-271.

\bibitem{GibRai} J. Gibbons and A. Raimondo,  Differential Geometry of Hydrodynamic Vlasov Equations, arXiv:nlin.SI/0612022.


\bibitem{Haantjes} J.Ê Haantjes, On $X\sb m$-forming sets of eigenvectors, Indagationes Mathematicae {\bf 17}  (1955) 158-162.

\bibitem{K} K.R. Khusnutdinova and S.P. Tsarev, private communication.


\bibitem{Kon2} B.G. Konopelchenko and L. Martinez Alonso, 
Dispersionless scalar integrable hierarchies, Whitham hierarchy, and 
the quasiclassical $\overline\partial$-dressing method, J. Math. 
Phys. {\bf 43}, no. 7 (2002)  3807-3823. 

\bibitem{Kon3} B.G. Konopelchenko, L. Martinez Alonso and O. 
Ragnisco, The $\overline\partial$-approach to the dispersionless KP 
hierarchy, J. Phys. A {\bf 34}, no. 47 (2001) 10209-10217. 

\bibitem{Krichever} I.M.Ê Krichever,  The $\tau$-function of the universal Whitham hierarchy, matrix models and topological field theories, Comm. Pure Appl. Math. {\bf 47}, no. 4 (1994) 437-475. 


\bibitem{Kup1} B.A. Kupershmidt and Yu.I. Manin, Long wave equations with a free surface. I. Conservation laws and solutions. (Russian) Funktsional. Anal. i Prilozhen. {\bf 11}, no. 3 (1977) 31-42.


\bibitem{Kup2} B.A. Kupershmidt and Yu.I. Manin, Long wave equations with a free surface. II. The Hamiltonian structure and the higher equations. (Russian) Funktsional. Anal. i Prilozhen. {\bf 12}, no. 1 (1978) 25-37.

\bibitem{Kup3} B.A. Kupershmidt,  Deformations of integrable systems. Proc. Roy. Irish Acad. Sect. A {\bf 83}, no. 1 (1983) 45-74.

\bibitem{Kup4} B.A. Kupershmidt, The quasiclassical limit of the modified KP hierarchy, J. Phys. A: Math. Gen {\bf 23} (1990) 871-886.

\bibitem{Kup5} B.A. Kupershmidt, Normal and universal forms in integrable hydrodynamical systems, 
Proceedings of the Berkeley-Ames conference on nonlinear problems in control and fluid dynamics (Berkeley, Calif., 1983), in Lie Groups: Hist., Frontiers and Appl. Ser. B: Systems Inform. Control, II, Math Sci Press, Brookline, MA, (1984) 357-378.

\bibitem{Kup6} B.A. Kupershmidt, Extended equations of long waves, preprint, 2005.

\bibitem{Lebedev} D.R. Lebedev and Yu.I. Manin, Conservation laws and Lax representation for Benney's long wave equations, Phys. lett. {\bf 74}A (1979) 154-156.

\bibitem{Ma1} M. Ma$\tilde {\rm n}$as, On the $r$th dispersionless Toda hierarchy: factorization problem, additional symmetries and some solutions, J. Phys. A: Math. Gen. {\bf 37} (2004) 9195-9224.

\bibitem{Ma2} M. Ma$\tilde {\rm n}$as, $S$-functions, reductions and hodograph solutions of the $r$th dispersionless modified KP and Dym hierarchies, J. Phys. A: Math. Gen. {\bf 37} (2004) 11191-11221.

\bibitem{Ma3} M. Ma$\tilde {\rm n}$as,  L. Martinez  Alonso and E. Medina,
Reductions and hodograph solutions of the dispersionless KP
hierarchy, J. Phys. A: Math. Gen. {\bf 35} (2002) 401-417.




\bibitem{Mikhalev} V.G. Mikhalev, Hamiltonian formalism of Korteweg-de Vries-type hierarchies. (Russian) Funktsional. Anal. i Prilozhen. {\bf 26}, no. 2 (1992) 79-82; translation in Funct. Anal. Appl. {\bf 26}, no. 2 (1992) 140-142.

\bibitem{Miura} R.M. Miura, Conservation laws for fully nonlinear long wave equations, Stud. Appl. Math.
{\bf 53} (1974) 45-56. 

\bibitem{Pavlov1} M.V. Pavlov, Integrable hydrodynamic chains,  J. 
Math. Phys.  {\bf 44}, no. 9 (2003) 4134-4156.

\bibitem{Pavlov2} M.V. Pavlov, Classification of  integrable
Egorov hydrodynamic chains, Teoret. Mat. Fiz. {\bf 138}, no. 1 (2004) 55-70.

\bibitem{Pavlov3} M.V. Pavlov, S.I.  Svinolupov and R.A.  Sharipov,  An invariant criterion for hydrodynamic integrability, Funktsional. Anal. i Prilozhen. {\bf 30} (1996) 18-29; translation in Funct. Anal. Appl. {\bf 30} (1996) 15-22.

\bibitem{Pavlov4} M.V. Pavlov, private communication, 2005.

\bibitem{Pavlov5} M.V. Pavlov, Modified dispersionless Veselov--Novikov equations and corresponding hydrodynamic chains, arXiv:nlin.SI/0611022.

\bibitem{Pavlov6} M.V. Pavlov,  The Hamiltonian approach in classification and integrability of hydrodynamic chains, arXiv:nlin.SI/0603057.


   
\bibitem{Perad2}Z. Peradzy\'nski, Nonlinear plane $k$-waves and
Riemann invariants, Bull. Acad. Polon. Sci. Sr. Sci. Tech. {\bf 19}
(1971) 625-632.


\bibitem{Sevennec} B. S\'evennec, G\'eom\'etrie des syst\`emes hyperboliques
de lois de conservation, M\'emoire (nouvelle s\'erie) N56,
Suppl\'ement au Bulletin de la Soci\'et\'e Math\'ematique de France {\bf 122} (1994) 1-125.

\bibitem{Shabat1} L. Martinez Alonso and A.B. Shabat,  Towards a theory of differential constraints of a hydrodynamic hierarchy, J. Nonlinear Math. Phys. {\bf 10}, no. 2  (2003) 229-242. 

\bibitem{Shabat2} L. Martinez Alonso and A.B. Shabat, 
Energy-dependent potentials revisited: A universal hierarchy of 
hydrodynamic type,
Phys. lett. A {\bf 299}, no. 4 (2002) 359-365;  Phys. Lett. A {\bf 300}, no. 1 (2002) 58-64.

\bibitem{T1} V.M. Teshukov,  On the hyperbolicity of long wave equations, Dokl. Akad. Nauk SSSR {\bf 284}, no. 3 (1985) 555-559.

\bibitem{T2} V.M. Teshukov, Characteristics, conservation laws and symmetries of the kinetic equations of motion of bubbles in a fluid,  Prikl. Mekh. Tekhn. Fiz. {\bf 40}, no. 2 (1999) 86-100; translation in J. Appl. Mech. Tech. Phys. {\bf 40}, no. 2 (1999) 263-275.


\bibitem{Tsarev} S.P. Tsarev, Geometry of Hamiltonian systems of
hydrodynamic type. Generalized hodograph method, Izvestija AN USSR
Math. {\bf 54}, no. 5  (1990) 1048-1068.

\bibitem{Zakharov1} V.E. Zakharov, Benney equations and quasiclassical approximation in the inverse problem method, Funktsional. Anal. i Prilozhen. {\bf 14}, no. 2 (1980) 15-24. 

\bibitem{Zakharov2} V.E. Zakharov, On the Benney equations, Physica {\bf 3}D, no. 1,2 (1981) 193-202.



\end{thebibliography}
\end{document}